\documentclass[journal]{IEEEtran}
\usepackage{amsmath,amssymb,amsfonts}
\usepackage{mathtools}
\usepackage{bm}
\usepackage{booktabs}
\usepackage{multirow}
\usepackage{array}
\usepackage{tabularx}
\usepackage{makecell}
\usepackage{graphicx}
\usepackage{algorithm}
\usepackage{algpseudocode}
\usepackage{textcomp}
\usepackage{listings}
\usepackage{url}
\usepackage{cite}
\usepackage{balance}
\usepackage{xcolor}
\usepackage{tcolorbox}
\tcbuselibrary{breakable,skins}
\usepackage[hidelinks]{hyperref}

\hypersetup{
  colorlinks=true,
  linkcolor=blue,
  citecolor=blue,
  urlcolor=blue,
  filecolor=blue
}

\newcolumntype{L}{>{\raggedright\arraybackslash}X}

\newcommand{\ok}[1]{\textcolor{green!50!black}{#1}}

\newcommand{\bad}[1]{\textcolor{red!70!black}{#1}}

\newcommand{\gov}[1]{\textcolor{purple!70!black}{#1}}
\newcommand{\sep}{\textcolor{black!35}{\rule{\textwidth}{0.35pt}}}

\providecommand{\Ok}[1]{\ok{#1}}

\providecommand{\Bad}[1]{\bad{#1}}

\providecommand{\Gov}[1]{\gov{#1}}
\providecommand{\Sep}{\par\noindent\sep\par}
\providecommand{\Hdr}[1]{\textbf{#1}}
\providecommand{\Dim}[1]{\textcolor{black!70}{#1}}
\providecommand{\Blk}[1]{\textbf{#1}}
\providecommand{\LogLine}[1]{#1\\}

\begin{document}

\title{Multi-Agent LLM Governance for Safe Two-Timescale Reinforcement Learning in SDN-IoT Defense}

\author{
Saeid~Jamshidi,
Negar Shahabi,
Foutse~Khomh,
Carol~Fung,
and Mohammad~Hamdaqa
\thanks{
S. Jamshidi and F. Khomh are with the SWAT Laboratory,
Polytechnique Montréal, Montréal, QC, Canada
(e-mail: \{saeid.jamshidi, foutse.khomh\}@polymtl.ca).
}
\thanks{
M. A. Hamdaqa is with the SæT Laboratory,
Polytechnique Montréal, Montréal, QC, Canada.
}
\thanks{
N. Shahabi and C. Fung are with the Concordia Institute for Information Systems Engineering (CIISE),
Concordia University, Montréal, QC, Canada
(e-mail: negar.shahabi@concordia.ca, carol.fung@concordia.ca).
}
}

\maketitle

\begin{abstract}
Software-Defined Networking (SDN) is increasingly adopted to secure Internet of Things (IoT) networks due to its centralized control and programmable forwarding. However, SDN-IoT defense is inherently a closed-loop control problem in which mitigation actions impact controller workload, queue dynamics, rule-installation delay, and future traffic observations. Aggressive mitigation may destabilize the control plane, degrade Quality of Service (QoS), and amplify systemic risk. Existing learning-based approaches prioritize detection accuracy while neglecting to model controller coupling and optimize short-horizon Reinforcement Learning (RL) objectives without structured, auditable policy evolution. This paper introduces a two-timescale, governance-driven SDN-IoT defense solution that separates fast, decentralized mitigation from slow policy governance. At the fast timescale, per-switch Proximal Policy Optimization (PPO) agents perform controller-aware mitigation under explicit safety constraints and action masking. On a slow timescale, a multi-agent Large Language Model (LLM) governance engine generates machine-parsable edits to a global policy constitution $\Pi$, thereby encoding admissible action masks, controller-aware safety thresholds, and reward priorities. Policy updates are formalized as structured deltas $\Delta\Pi$, validated through targeted stress campaigns, and deployed only if non-regression and hard-safety conditions are met, ensuring fail-closed, auditable evolution without modifying RL parameters. Evaluation is conducted as a closed-loop systems study under heterogeneous IoT traffic and adversarial stress, explicitly modeling controller service-rate limits and delayed actuation. Results show that the proposed solution improves Macro-F1 by 9.1\% over unconstrained PPO and by 15.4\% over static threshold-based mitigation. Worst-case agent degradation is reduced by 36.8\%, controller backlog peaks decrease by 42.7\%, and RTT p95 inflation remains below 5.8\% during high-intensity attacks. No oscillatory instability and controller collapse are observed across independent runs. Governance-driven policy evolution converges within five reflection cycles, reducing catastrophic overload events from 11.6\% to 2.3\% of episodes without retraining RL parameters.
\end{abstract}


\section{Introduction}
\label{sec:introduction}
The rapid proliferation of Internet of Things (IoT) devices has transformed enterprise and critical infrastructures into large-scale, heterogeneous, and latency-sensitive environments \cite{dritsas2025survey}\cite{cheikh2026energy}. Industrial systems, smart grids, transportation networks, and healthcare platforms now rely on billions of resource-constrained devices that continuously generate dynamic traffic  \cite{hudda2025review}\cite{ayyadurai2026cloud}. This expansion substantially increases the attack surface, exposing operational networks to Denial-of-Service (DoS) attacks, stealthy scans, and adaptive adversarial behaviors \cite{mishra2025sdniotreview, rahdari2024sdniot}.    
Software-Defined Networking (SDN) is widely adopted in IoT networks due to its centralized control plane and programmable forwarding capabilities \cite{chaudhary2025comprehensive}. SDN enables rapid deployment of mitigation actions, e.g., flow drops, rate limiting, and quarantine policies \cite{mulugetareview}. However, this centralization introduces structural coupling: mitigation actions themselves consume controller resources and alter queue dynamics, flow-table states, and rule-install delays. Consequently, defensive decisions modify not only the attack surface but also the distribution of future observations \cite{mcmahan2024optimal}\cite{fang2024adaptive}. SDN-IoT defense, therefore, constitutes a closed-loop control problem rather than a pure traffic classification task \cite{mustafa2024intrusion}\cite{mozumder2025smartsecchain}. 
Aggressive mitigation actions in SDN networks may increase control-plane overhead, as they often require frequent rule installations and interactions with the controller, which can significantly impact controller workload and network performance \cite{troia2025comprehensive}\cite{bhuiyan2023security}\cite{razvan2025enhancing}. Under heavy traffic and attack conditions, this may lead to controller bottlenecks and degradation of Quality of Service (QoS) \cite{khozam2025qosentry,aguirre2025qosdrl}. Despite these operational constraints, many existing security mechanisms treat mitigation as an isolated response step rather than as part of a coupled control process involving the controller, switches, and network dynamics. Existing learning-based SDN security solutions generally follow two paradigms. Detection-centric approaches primarily focus on improving classification accuracy, treating mitigation as a static reaction layer, and often neglecting the feedback between mitigation decisions and control-plane dynamics \cite{kalambe2025comprehensive,hozouri2025comprehensive,khanal2025real}. Reinforcement learning (RL) approaches have recently been explored to enable adaptive mitigation and dynamic network control in SDN networks \cite{bijlwan2026deep,balasubramanian2025iot}. However, many RL-based systems optimize simplified reward functions and short-term objectives, which may lead to unstable behavior in the presence of complex traffic dynamics and adversarial traffic patterns \cite{michailidis2025traffic,skoropad2025dynamic}. Furthermore, most existing systems lack mechanisms for structured and auditable policy evolution. When instability occurs, adaptation typically relies on retraining models and manually adjusting mitigation rules, which can be operationally costly and difficult to audit in large-scale SDN-IoT networks \cite{buczak2016survey}. As a result, current approaches struggle to simultaneously maintain adaptive defense behavior, controller stability, and QoS guarantees under evolving attack conditions. To address these limitations, we introduce a two-timescale control framework for SDN-IoT defense. On fast timescales, decentralized per-switch Proximal Policy Optimization (PPO) \cite{qiu2026plasticity} agents perform controller-aware mitigation under explicit safety constraints. At the slow timescale, a governance layer analyzes system trajectories and proposes structured updates to a shared global policy entity $\Pi$. Instead of modifying neural network parameters, this governance mechanism updates an explicit policy constitution that defines admissible action masks, controller-aware thresholds, reward priorities, and actuation limits. Policy changes are represented as machine-readable updates, $\Delta\Pi$, and are deployed only after passing non-regression tests and safety validation, ensuring fail-safe operation. By separating fast mitigation from slower policy evolution, the proposed design combines adaptive learning with conservative network change management. This two-timescale structure preserves decentralized responsiveness while limiting high-risk behavior caused by controller coupling. The main contributions of this work are:
\begin{itemize}
\item \textbf{two-timescale governance-driven two-timescale SDN-IoT defense architecture.}
We formulate SDN-IoT defense as a closed-loop control system that separates fast decentralized mitigation from slower LLM governance. Independent per-switch PPO agents perform real-time mitigation using deployable telemetry signals, while a multi-agent LLM governance layer operates on a slower timescale to analyze trajectory evidence and adapt the global safety structure. This separation preserves rapid local responsiveness while enabling stable and interpretable long-horizon policy evolution.
\item \textbf{Auditable policy constitution with structured multi-agent LLM updates.}
We introduce a global policy entity, $\Pi$, that serves as an explicit policy constitution encoding safety constraints, admissible action sets, controller-aware thresholds, and reward priorities. Policy evolution is expressed through structured updates ($\Delta\Pi$) generated by a constrained multi-agent LLM pipeline (Critic, Compiler, Red-Team, and Judge). Updates are schema-validated, conflict-resolved, and deployed only after deterministic validation and non-regression testing, ensuring interpretable, auditable, and reproducible policy evolution.
\item \textbf{Controller-aware and risk-sensitive defense under realistic SDN dynamics.}
The framework explicitly models SDN control-plane coupling, including controller backlog, delayed rule activation, and flow-table pressure, capturing how mitigation actions reshape future network states. Defense policies are optimized using a risk-sensitive, multi-objective formulation that jointly balances security effectiveness, QoS stability, controller load, and operational overhead, thereby encouraging policies that prevent catastrophic controller saturation while maintaining stable network operation.
\end{itemize}

The remainder of this paper is organized as follows. Section~\ref{sec:related} reviews related work on SDN-IoT security. Section~\ref{sec:method} presents the proposed two-timescale governance-driven SDN-IoT control framework. Section~\ref{sec:threat} defines the threat model. Section~\ref{sec:experimental_setup} describes the experimental setup and evaluation protocol. Section~\ref{System-Level Performance Evaluation} reports system-level performance results. Section~\ref{sec:comparison} compares the proposed approach with prior work. Section~\ref{sec:limitations} discusses limitations and future work. Section~\ref{sec:conclusion} concludes the paper.

\section{Related Work}
\label{sec:related}
This section reviews prior research on SDN-IoT security, focusing on intrusion detection, autonomous defense, multi-agent architectures, and resilient explainable security mechanisms.

\subsection{Deep-Learning-Based Security in SDN-IoT Networks}
A substantial body of recent work focuses on DL-driven intrusion detection for SDN-enabled IoT networks. These studies exploit the centralized visibility of SDN controllers and the programmability of SDN-IoT architectures to detect attacks across traffic flows, packet-level attributes, and temporal communication patterns. For example, the HCL framework in \cite{khalis2026robust} combines CNN and LSTM models to capture both spatial and temporal traffic characteristics and reports effective detection of DoS, DDoS, port scanning, fuzzing, and OS-level attacks in SDN-IoT settings. The study also highlights persistent limitations in existing SDN-IoT IDS research, including scalability concerns, challenges in multiclass classification, and high computational demands. Other works similarly emphasize that DL improves detection capabilities in SDN-IoT environments but often focus primarily on classification performance. The secure routing framework in \cite{kokila2025deepsdn} integrates deep-belief-network-based attack detection with QoS-aware routing, thereby combining security and routing efficiency in SDN-IoT networks. The DeepSDN architecture in \cite{chouhan2025hcl} advances controller-centric security orchestration by coupling DL with SDN-based cloud supervision and blockchain-based communication, enabling real-time detection and mitigation of multiple classes of cyber threats in IoT networks. Likewise, the robust CNN-based multiclass IDS in \cite{shi2025autonomous} demonstrates that CNN-based models can generalize across multiple SDN-IoT datasets while maintaining strong performance under class imbalance and in multiclass attack settings.

\subsection{Reinforcement Learning for Autonomous Cyber Defense}
A second line of work shifts from passive detection to adaptive defense through reinforcement learning. These studies treat cyber defense as a sequential decision-making problem, allowing agents to learn policies through interaction with simulated and operational environments. In \cite{gan2025adaptive}, graph-attention-enhanced DRL is proposed for autonomous AIoT defense, leveraging graph representations to encode relationships among hosts and interactions between devices. This work argues that one-dimensional observations are insufficient for complex IoT topologies and shows that graph-aware PPO and DQN variants can improve autonomous defense performance in increasingly complex scenarios. A closely related direction is represented by \cite{wang2025novel}, which enhances autonomous cyber-defense decision-making through graph embedding. That work argues that traditional one-hot state encodings fail to capture topology and inter-node dependence, and therefore proposes Node2vec-enhanced RL over a game-theoretic attacker–defender setting. Importantly, \cite{wang2025novel} also highlights two limitations highly relevant to the present study: many RL-based cyber-defense methods neglect topology-aware perception, and many available training environments remain overly abstract and insufficiently representative of real operational settings. 
Recent research has also explored autonomous cyber-defense frameworks that combine RL with additional AI components. In \cite{loevenich2025design}, a DRL-based autonomous cyber-defense agent is implemented together with knowledge graphs and an augmented LLM, enabling monitoring, service removal, recovery, and decoy deployment in a simulated cyber-operations environment. This work is particularly relevant because it notes that earlier DRL-based approaches often behave as black boxes and lack explicit network-topology representation and structured human–machine interaction. In addition, \cite{cadet2025quantitative} introduces resilience-oriented PPO-based defensive agents and argues that autonomous defenses should be evaluated not only by cumulative reward but also by their ability to absorb attacks, recover promptly, and preserve defender priorities over time.

\subsection{Multi-Agent and Distributed Defense Architectures}
A third category of work investigates multi-agent and distributed cyber-defense architectures. These studies are motivated by the limitations of centralized security systems in large-scale, dynamic, and distributed environments. In \cite{lucia2026cyberspade}, a hierarchical multi-agent architecture is proposed for coordinated cyber defense using specialized swarms and explicit inter-agent communication. The work emphasizes that effective cyber defense increasingly requires cooperation, communication, and task allocation among distributed agents rather than isolated detection modules. Similarly, \cite{selvaraj2025cognitive} presents a rule-based cognitive AI framework in which multiple agents specialize in threat detection, risk assessment, mitigation planning, and adaptive learning, thereby improving interpretability and response coordination. The multi-agent perspective is also evident in \cite{aydin2025multi}, which proposes a cloud-oriented agile defense system for volumetric DDoS mitigation that leverages LSTM-based detection, federated learning, blockchain-backed transparency, and multiple specialized agents. This work highlights the operational advantages of agent collaboration, including parallel execution, reduced latency, local learning, and scalability across large environments. Although its application domain is public-cloud defense rather than SDN-IoT, it demonstrates how distributed agents can support autonomous detection and mitigation while preserving privacy and system responsiveness. More control-theoretic perspectives on distributed secure coordination appear in \cite{xu2025distributed}, which studies resilient control of cyber-physical multi-agent systems under attacks. While \cite{xu2025distributed} does not specifically target SDN-IoT intrusion detection, it formalizes the use of trust, confidence, attack localization, and resilient distributed control to maintain secure collective behavior under adversarial conditions. From a system design perspective, this reinforces the potential advantages of distributed defensive intelligence over purely centralized systems.

Although previous work has made significant strides in improving detection accuracy, adaptive mitigation, and scalable multi-agent control in SDN-IoT networks, these approaches often rely on a single optimization loop in which policies evolve primarily through parameter updates. As a result, the intricate structural coupling between mitigation actions, controller workload, delayed actuation, and QoS stability remains largely unexplored. Additionally, existing systems lack mechanisms for auditable and conservative policy evolution. When instability and unexpected behavior arise, adaptation typically depends on retraining and manual rule tuning, both of which are opaque and operationally costly. These limitations highlight the need to reframe SDN-IoT defense as a two-timescale control challenge, in which fast, decentralized mitigation operates alongside slower, structured governance that enables the evolution of policy constraints without destabilizing learned behavior. The solution proposed in this paper aims to bridge this gap by integrating controller-aware reinforcement learning with interpretable policy-constitution editing validated through stringent safety checks.

\section{Proposed Two-Timescale Governance-Driven Multi-Agent SDN-IoT Control Solution}
\label{sec:method}
\begin{figure*}[t]
  \centering
  \includegraphics[width=0.80\textwidth]{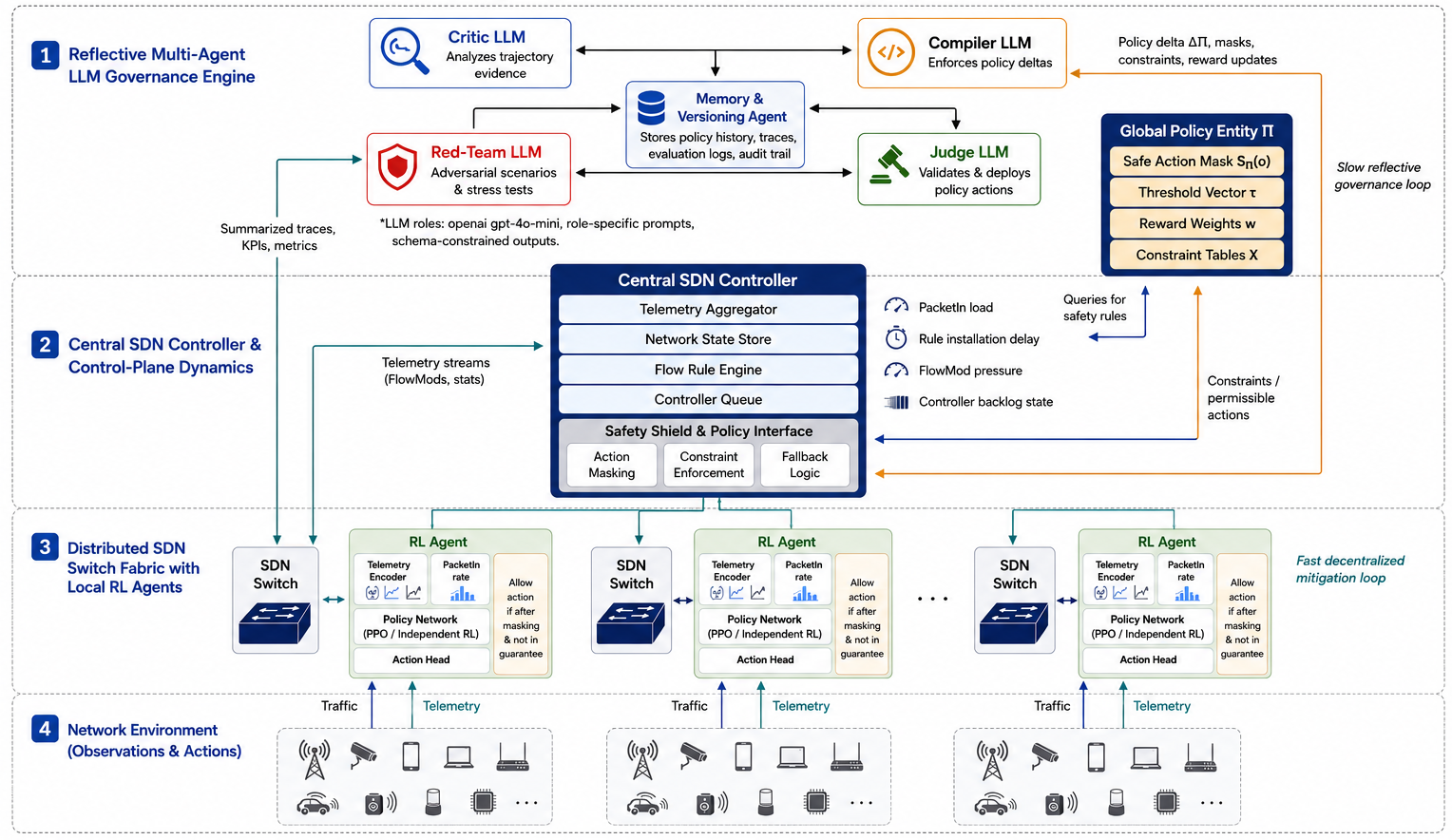}
  \caption{Architecture of the proposed two-timescale SDN-IoT defense solution. Independent per-switch PPO agents provide fast local mitigation, the SDN controller enforces controller-aware safety constraints, and the LLM governance layer operates on a slower timescale to update the global policy constitution $\Pi$. The LLM roles generate, test, and approve structured policy deltas $\Delta\Pi$ rather than directly controlling packet-level actions and modifying PPO parameters.}
  \label{fig:sdn_iot_reflective_framework}
\end{figure*}

The proposed solution models SDN-IoT defense as a closed-loop control problem in which mitigation decisions impact packet forwarding, controller workload, rule-installation delay, flow-table pressure, and future observations. The architecture has three layers. The fast layer contains an independent structure as Multi-agent RL; rather, it consists of actions from deployable SDN primitives. The middle layer is the SDN controller and safety-filter interface, which aggregates telemetry, mediates shared control-plane dynamics, and enforces controller-aware action constraints. The slow layer is the LLM governance layer, which neither selects packet-level actions nor modifies PPO parameters. Instead, it periodically analyzes summarized trajectory evidence and proposes structured edits to a global policy constitution $\Pi$. In this paper, $\Pi$ denotes the explicit set of safety masks, controller thresholds, reward priorities, and actuation limits that constrain the PPO agents. A policy update is denoted by $\Delta\Pi$ and is deployed only after schema validation, stress testing, hard-safety checks, and non-regression evaluation. Thus, PPO agents answer ``what action should this switch take now?'', whereas the LLM governance layer answers ``how should the global safety rules be adjusted for future decisions?'' Figure~\ref{fig:sdn_iot_reflective_framework} illustrates this two-timescale architecture.
The term multi-agent is used in two distinct senses. First, the SDN environment contains multiple switch-level RL agents that interact indirectly through shared controller backlog and flow-table resources. These agents are trained independently for each switch; therefore, the learning layer is not formulated as centralized multi-agent RL and does not use a centralized critic, joint gradient sharing, and a joint MARL objective. Second, the LLM governance layer contains multiple functional roles, namely Critic, Compiler, Red-Team, Judge, and Memory, which diagnose failures, compile candidate policy edits, generate stress campaigns, and approve updates. These LLM roles are governance roles, not data plane control agents. Decentralized PPO provides fast local responsiveness, while global coordination is induced through shared controller dynamics and the $\Pi$-constrained action space and reward shaping.
For example, if recent traces show that frequent \texttt{DROP\_FLOW} and \texttt{QUARANTINE} actions increase controller backlog and delay rule activation, the LLM governance layer may propose a $\Delta\Pi$ that temporarily masks controller-heavy actions when $d^t$ exceeds a predefined threshold, increases the controller-stress penalty in the reward, and requires validation under burst, mimicry, and multi-switch stress campaigns before deployment. If the candidate update fails validation, the system retains the previous policy constitution $\Pi$ and continues without modification. This example illustrates that the LLM layer governs the evolution of safety constraints and reward priorities rather than directly controlling packets, switches, and PPO parameters.
We represent the latent global network state as:
\begin{equation}
s^t =
\big(
\mathbf{q}^t,\;
\mathbf{f}^t,\;
\mathbf{d}^t,\;
\mathbf{z}^t
\big),
\end{equation}
where $\mathbf{q}^t \in \mathbb{R}_+^{N}$ denotes switch queue occupancies, $\mathbf{f}^t$ represents flow-table states including timeout counters and eviction risk, $\mathbf{d}^t$ captures controller backlog, service delay, and pending rule installations, and $\mathbf{z}^t$ encodes latent traffic, including benign load and adversarial intensity. This decomposition makes explicit that SDN-IoT defense is not a static classification task: mitigation changes queueing, rule availability, and control-plane latency, thereby altering future observations.
Formally, the SDN-IoT control loop defines a controlled Markov process with transition kernel:
\begin{equation}
P:\; (\mathcal{S}\times\mathcal{A}) \rightarrow \Delta(\mathcal{S}),
\end{equation}
where $\Delta(\mathcal{S})$ denotes the probability simplex over $\mathcal{S}$. The system evolves according to:
\begin{equation}
s^{t+1}\sim P(s^{t+1}|s^t,\mathbf{a}^t),
\end{equation}
where $\mathbf{a}^t=(a_1^t,\dots,a_N^t)$ is the joint action vector. Because controller capacity is shared, this transition kernel is non-separable: local decisions alter the transition distribution experienced by all agents. In SDN, delayed actuation means that controller queueing and rule-apply delay impact the interval between decision and effect.
To make rule-activation delay explicit, we introduce an auxiliary controller memory state $\xi^t$ that captures pending FlowMod and meter-install jobs:
\begin{equation}
s^{t+1} \sim P\!\left(s^{t+1}\mid s^t,\mathbf{a}^t,\xi^t\right),
\qquad
\xi^{t+1}=H\!\left(\xi^t,\mathbf{a}^t,d^t\right).
\end{equation}
Here, $\xi^t$ formalizes that \texttt{DROP\_FLOW} and \texttt{QUARANTINE} may not become effective immediately when controller backlog is high. Their effect is mediated through $\xi^t$ and $d^t$. We incorporate SDN control-plane realism into $P(\cdot)$ by modeling bounded PacketIn buffering, stochastic controller service time, stochastic rule-apply delay, and finite flow-table capacity with timeout and eviction dynamics. These mechanisms determine when an action becomes effective and how long flow misses persist.
The controller backlog evolves as:
\begin{equation}
d^{t+1} =
\max\!\left(
0,\,
d^t + \sum_{i=1}^{N}\rho_i^t - \mu
\right),
\end{equation}
where $d^t$ represents pending PacketIn and rule-installation requests at time $t$, $\rho_i^t$ is the PacketIn arrival rate generated by switch $i$, and $\mu$ is the controller service rate. This dynamic clarifies that even correct mitigation may destabilize the system if it induces excessive PacketIn and rule churn. When $\sum_i\rho_i^t$ approaches and exceeds $\mu$, the controller becomes the bottleneck, rule installation is delayed, flow misses persist longer, and observations become biased by control-plane saturation.
A standard sufficient condition for bounded backlog in expectation is:
\begin{equation}
\limsup_{T\to\infty}\frac{1}{T}\sum_{t=0}^{T-1}\mathbb{E}\!\left[\sum_{i=1}^N \rho_i^t\right] < \mu.
\end{equation}
This inequality motivates controller-aware shaping terms, e.g., $g_{\text{ctrl}}^t$ and $\Pi$-based caps. Policies that reduce attacks but increase $\sum_i\rho_i^t$ can violate this condition and destabilize the control plane. Although a full Lyapunov proof is beyond scope, the design induces a negative-feedback stabilization mechanism: when $d^t$ and related overload indicators increase, $\Pi$ tightens admissible actions and throttles controller-heavy actuation, thereby reducing effective PacketIn arrivals, rule churn, and delayed-actuation cascades.
Let $t$ denote the fast RL timestep and let reflection occur every $K$ steps. Define the slow index $k=\lfloor t/K\rfloor$ and the trace window $\mathcal{T}_k \equiv \mathcal{T}_{t-K:t}$. Governance evolution is:
\begin{equation}
\begin{aligned}
\Pi_{k+1} &= \Pi_k \oplus \Delta\Pi_k,\\
\Delta\Pi_k &= \mathcal{L}(\Pi_k,\mathcal{E}_k),\\
\mathcal{E}_k &= \textsc{SummarizeTrace}(\mathcal{T}_k).
\end{aligned}
\end{equation}
It is important to distinguish $\Pi$ from the PPO policies $\{\pi_{\theta_i}\}$. The local PPO policy $\pi_{\theta_i}(a_i|o_i)$ maps switch-level observations to mitigation actions. In contrast, $\Pi$ is not an RL policy and does not directly select actions. Instead, $\Pi$ defines the effective operating environment in which the PPO agents act. It determines which actions are admissible, which controller conditions trigger safety constraints, how reward components are weighted, and which actuation limits are enforced. Therefore, the LLM governance layer modifies the environment-facing safety and reward structure around the PPO agents, rather than replacing and retraining the PPO policies.
We define:
\begin{equation}
\Pi \triangleq 
\big(
\mathcal{M}_{\Pi}(\cdot),\;
\boldsymbol{\tau},\;
\mathbf{w},\;
\boldsymbol{\chi}
\big),
\end{equation}
where $\mathcal{M}_{\Pi}(o)$ denotes the context-dependent action-mask function, $\boldsymbol{\tau}$ denotes controller-aware safety thresholds, $\mathbf{w}$ denotes reward-shaping weights used in $R^t=\mathbf{w}^{\top}\mathbf{g}^t$, and $\boldsymbol{\chi}$ denotes auxiliary governance patches, e.g., action-dependent cost tables, rule-install throttles, controller-aware actuation caps, and context-specific fallback rules.
Under this interpretation, $\Pi$ does not conflate heterogeneous elements into an RL policy. Rather, it groups the environment-facing governance mechanisms that shape how PPO actions are filtered, penalized, delayed, and deployed. Changing $\boldsymbol{\tau}$ alters when controller overload constraints become active; changing $\mathcal{M}_{\Pi}$ changes the feasible action set; changing $\mathbf{w}$ reshapes the optimization signal; and changing $\boldsymbol{\chi}$ modifies execution rules, e.g., throttling and fallback behavior.
The LLM governance layer updates $\Pi$ by producing a machine-parsable delta $\Delta\Pi_k$ from trace evidence:
\begin{equation}
\Delta\Pi_k = \mathcal{L}(\Pi_k,\mathcal{E}_k),
\end{equation}
where $\Delta\Pi_k$ is restricted to edits of 
$(\mathcal{M}_{\Pi},\boldsymbol{\tau},\mathbf{w},\boldsymbol{\chi})$
and is explicitly forbidden from editing PPO weights $\{\theta_i\}$. The allowed edit set is intentionally conservative: $\Delta\Pi_k$ may only restrict and relax masks within predefined action families, adjust thresholds within bounded ranges, and reweight shaping terms under predefined context flags. The evidence bundle $\mathcal{E}_k=\textsc{SummarizeTrace}(\mathcal{T}_k)$ is computed deterministically from control-plane counters, dataplane counters, and available evaluation labels. It includes queue tails, FlowMod rates, PacketIn drops, RTT percentiles, churn, false positives, and false negatives. The LLM roles are not given raw packet payloads and do not require deep packet inspection.

\subsection{Multi-Agent Model}
Each switch is associated with an autonomous decision agent, yielding a partially observable stochastic game:
\begin{equation}
\mathcal{G}=
\langle
\mathcal{S},
\{\mathcal{O}_i\},
\{\mathcal{A}_i\},
P,
R,
\gamma
\rangle.
\end{equation}
Here, $\mathcal{S}$ is the latent global SDN-IoT state space, $\mathcal{O}_i$ is the local observation space of switch agent $i$, $\mathcal{A}_i$ is its discrete mitigation-action space, $P$ is the controlled state-transition kernel, $R$ is the shared reward signal, and $\gamma$ is the discount factor. The formulation is partially observable because each switch observes only local telemetry, while the full network state includes shared controller backlog, flow-table pressure, delayed rule activation, and latent benign and adversarial traffic conditions.
At each time $t$, agent $i$ receives telemetry:
\begin{equation}
o_i^t=
\{r_i^t,q_i^t,c_i^t,f_i^t,e_i^t,u_i^t,\rho_i^t,\varphi_i^t,h_i^t\},
\end{equation}
where $r_i^t$ denotes traffic rate, $q_i^t$ denotes queue occupancy, $c_i^t$ denotes local processing pressure, $f_i^t$ denotes flow-table pressure, $e_i^t$ denotes normalized traffic-source entropy, $u_i^t$ denotes port and destination diversity, $\rho_i^t$ denotes PacketIn intensity generated by switch $i$, $\varphi_i^t$ denotes recent actuation intensity, e.g., FlowMod and meter-install frequency, and $h_i^t$ denotes weak contextual hints and context flags derived from telemetry summaries. PacketIn refers to the OpenFlow event in which a switch sends packet metadata to the controller when no matching forwarding rule is available. FlowMod refers to a controller instruction that adds, modifies, and removes forwarding rules, whereas meter-install refers to installing rate-limiting rules. These variables are computable from SDN/IoT monitoring primitives, e.g., per-port counters, queue occupancy, flow-table counters, PacketIn events, and FlowMod counters, without inspecting packet payloads.
The telemetry is a noisy projection of the latent global state, captured by:
\begin{equation}
o_i^t \sim O_i(\cdot \mid s^t),
\qquad
\mathbf{o}^t \sim \prod_{i=1}^N O_i(\cdot \mid s^t),
\end{equation}
which emphasizes that agents act on deployable counters rather than the full latent state. Entropy is computed as:
\begin{equation}
e_i^t=
-\frac{1}{\log K}
\sum_{k=1}^{K}p_{i,k}^t\log p_{i,k}^t,
\end{equation}
where $p_{i,k}^t$ is the empirical fraction of traffic associated with source and destination bin $k$ at switch $i$ and time $t$, and $K$ is the number of bins.
This term measures traffic-source dispersion and provides a scale-invariant indicator of distributed attacks. Because IoT traffic may include benign synchronized bursts, $e_i^t$ is interpreted jointly with $\rho_i^t$ as control-plane stress, with $\varphi_i^t$ as actuation intensity, and with $u_i^t$ as port and destination diversity, thereby reducing false alarms from legitimate events, e.g., firmware updates.
The multi-agent structure is used in two distinct senses. First, the RL layer contains $N$ switch-level agents that make local mitigation decisions and interact indirectly via shared SDN resources, particularly the controller backlog and flow-table capacity. These agents are trained independently; therefore, the method is not centralized multi-agent RL and does not rely on a centralized critic, joint gradients, and a joint MARL objective. Second, the LLM governance layer contains multiple governance roles that operate on aggregated trace evidence. These roles do not execute dataplane actions. Instead, they propose, compile, test, and gate updates to the governance configuration $\Pi$, thereby shaping the operating envelope of the PPO agents through the action-mask function $\mathcal{M}_{\Pi}(\cdot)$, controller-aware thresholds, reward-shaping weights, and auxiliary execution rules.
Operationally, switch-level PPO agents answer ``what action should be taken now at this switch given local telemetry?'', whereas LLM governance roles answer ``how should shared safety constraints, thresholds, and reward priorities be adjusted for the next operating horizon?'' This separation keeps the RL layer lightweight and reactive, while allowing the governance layer to perform slower policy configuration updates using aggregated evidence.
We model each LLM governance role as a typed mapping over evidence $\mathcal{E}_k$:
\begin{align}
\textsc{CriticLLM}:&
\;(\Pi_k,\mathcal{E}_k)\!\rightarrow\!\mathcal{I}_k
&&\text{(diagnosis)},\\
\textsc{CompilerLLM}:&
\;\mathcal{I}_k\!\rightarrow\!\Delta\Pi_k
&&\text{(compile)},\\
\textsc{RedTeamLLM}:&
\;\mathcal{E}_k\!\rightarrow\!\mathcal{C}_{rt}
&&\text{(stress tests)},\\
\textsc{JudgeLLM}:&
\;(\Pi_k,\Pi_{cand},\Delta\mathcal{M})
\!\rightarrow\!\{0,1\}
&&\text{(gate)}.
\end{align}
The Critic identifies failure modes, e.g., controller saturation, rule-churn bursts, and false disruptions under benign synchronized events. The Compiler converts the diagnosis into a candidate delta $\Delta\Pi_k$ containing action-mask updates, threshold changes, reward-weight adjustments, and actuation caps. The Red-Team generates targeted stress campaigns $\mathcal{C}_{rt}$ to evaluate whether the candidate governance configuration remains stable under adversarial and near-saturation conditions. The Judge produces an explicit approve/reject decision based on validation outcomes. Thus, the LLM governance layer is procedural and constrained: it produces structured updates to $\Pi$ and cannot directly select switch actions and modify PPO weights.
The LLM governance layer can be viewed as solving a conservative governance-configuration update:
\begin{equation}
\begin{aligned}
\Delta\Pi_k
&\in
\arg\max_{\Delta\Pi \in \mathcal{U}}
\;\mathbb{E}\!\left[\mathcal{J}(\Pi_k \oplus \Delta\Pi)\mid \mathcal{E}_k\right]\\
&\text{s.t.}\quad
\textsc{HardSafety}(\Pi_k \oplus \Delta\Pi)=1.
\end{aligned}
\end{equation}
Here, $\mathcal{U}$ denotes the bounded set of allowed edit types, including action-mask edits, threshold adjustments, reward-shaping updates, and actuation caps. The function $\textsc{HardSafety}(\cdot)$ is a deterministic validation function that checks controller-capacity limits, action-feasibility constraints, admissible threshold ranges, and internal consistency of the candidate governance configuration. This formulation explains why the LLM layer is not a free-form controller: it can only propose bounded updates to the environment-facing governance configuration, and those updates must satisfy hard-safety constraints before deployment.

\subsection{Coupled Action Effects in SDN}
\label{subsec:coupled_action_effects}
Each switch-level agent selects one mitigation action:
\begin{equation}
\begin{aligned}
a_i^t \in \{&
\texttt{ALLOW},\;
\texttt{ALERT},\;
\texttt{MIRROR},\\
&
\texttt{RATE\_LIMIT},\;
\texttt{DROP\_FLOW},\;
\texttt{QUARANTINE}
\}.
\end{aligned}
\end{equation}
These actions correspond to common SDN enforcement primitives. \texttt{ALLOW} forwards traffic without intervention, \texttt{ALERT} records suspicious behavior without changing forwarding behavior, and \texttt{MIRROR} duplicates selected traffic to a monitoring component. \texttt{RATE\_LIMIT} installs a meter rule that caps suspicious-flow transmission rate, \texttt{DROP\_FLOW} installs a forwarding rule that discards matching packets, and \texttt{QUARANTINE} redirects suspicious traffic to a restricted path, sink, and inspection segment. In OpenFlow-based SDN, controller-mediated actions, e.g., \texttt{RATE\_LIMIT}, \texttt{DROP\_FLOW}, and \texttt{QUARANTINE} may require FlowMod messages to add, modify, and remove switch forwarding rules. Meter-install operations install rate-limiting rules. These controller-mediated operations consume control-plane resources and may be delayed when the controller backlog is high.
These actions impact both data-plane traffic and control-plane dynamics, thereby altering future observations. For example, \texttt{DROP\_FLOW} and \texttt{QUARANTINE} can reduce malicious traffic but increase FlowMod load, while \texttt{MIRROR} can increase processing and bandwidth overhead. Therefore, the same action can yield different outcomes depending on controller saturation, rule-installation delay, and flow-table pressure.
Queue evolution can be written as:
\begin{equation}
q_i^{t+1}
=
\Big[
q_i^t
+
\lambda_i^t(a_i^t,\mathbf{z}^t)
-
\mu_i(a_i^t)
\Big]_+,
\end{equation}
where $\lambda_i^t(a_i^t,\mathbf{z}^t)$ denotes the effective offered load after mitigation under latent traffic condition $\mathbf{z}^t$, $\mu_i(a_i^t)$ denotes the effective service rate under action overhead, and $[\cdot]_+$ ensures non-negativity. For example, \texttt{RATE\_LIMIT} can reduce the offered load, whereas \texttt{MIRROR} can reduce the effective service due to monitoring overhead.
To model delayed actuation, we introduce a controller-induced delay-accumulation term rather than a reward penalty. This term represents additional queue growth caused by controller processing latency and delayed rule activation. Actions requiring controller intervention may not impact forwarding immediately when the controller backlog is high. Therefore, the queue update becomes:
\begin{equation}
q_i^{t+1}
=
\Big[
q_i^t
+
\lambda_i^t(a_i^t,\mathbf{z}^t)
-
\mu_i(a_i^t)
+
\delta_i^t(d^t,a_i^t)
\Big]_+,
\end{equation}
where $\delta_i^t(d^t,a_i^t)$ is the delay-induced queue accumulation associated with controller backlog $d^t$ and action $a_i^t$. This term is larger for controller-heavy actions, especially when the backlog is high.
A simple controller-aware model assumes monotone delay-induced accumulation:
\begin{equation}
\delta_i^t(d^t,a_i^t)= \kappa(a_i^t)\cdot \sigma(d^t),
\qquad
\sigma(d)=\log(1+d),
\end{equation}
where $\kappa(a_i^t)$ is an action-dependent delay coefficient. Larger values correspond to controller-heavy actions, e.g.,\texttt{DROP\_FLOW}, \texttt{QUARANTINE}, and \texttt{RATE\_LIMIT}; smaller values correspond to actions requiring little rule installation, e.g., \texttt{ALLOW}, \texttt{ALERT}, and \texttt{MIRROR}. This formalizes why the same action can have different effects under different controller-load operating conditions.
To make the delayed-actuation model reproducible, Table~\ref{tab:kappa_values} reports the $\kappa(a)$ values used in the experiments. These values encode the relative dependence of each SDN action on the controller. Actions with small $\kappa(a)$ introduce little controller-mediated delay, whereas actions with large $\kappa(a)$ require controller involvement through rule installation, flow-table modification, and quarantine handling.
\begin{table}[t]
\centering
\caption{Controller-dependence coefficient $\kappa(a)$ used in the delayed-actuation model.}
\label{tab:kappa_values}
\begin{tabular}{lcl}
\toprule
\textbf{Action} & \textbf{$\kappa(a)$} & \textbf{Interpretation} \\
\midrule
\texttt{ALLOW}       & 0.00 & No mitigation overhead \\
\texttt{ALERT}       & 0.05 & Logging/notification overhead \\
\texttt{MIRROR}      & 0.15 & Monitoring overhead \\
\texttt{RATE\_LIMIT} & 0.35 & Meter-rule installation \\
\texttt{DROP\_FLOW}  & 0.65 & Flow-rule installation/churn \\
\texttt{QUARANTINE}  & 0.90 & Highest controller overhead \\
\bottomrule
\end{tabular}
\end{table}
Thus, mitigation induces a coupled nonlinear control process:
\begin{equation}
(\mathbf{q}^{t+1},\mathbf{d}^{t+1})
=
F(\mathbf{q}^t,\mathbf{d}^t,\mathbf{a}^t,\mathbf{z}^t),
\end{equation}
where $F(\cdot)$ denotes the coupled SDN state-transition operator that aggregates queue evolution, controller-backlog dynamics, delayed rule activation, flow-table effects, and traffic-pattern changes induced by the joint action vector $\mathbf{a}^t$.
This formulation shows why purely centralized and purely heuristic policies can fail in large SDN-IoT networks: the same mitigation action can yield different outcomes depending on controller saturation, actuation delays, and flow-table pressure. This motivates the proposed two-timescale design: PPO agents handle immediate local variations, while the LLM governance layer updates the shared governance configuration that defines safe actions, controller-aware thresholds, and long-horizon priorities across switches.

\subsection{Risk-Sensitive Multi-Objective Reward}
We define the performance vector:
\begin{equation}
\mathbf{g}^t=
(g_{\text{sec}}^t,
g_{\text{lat}}^t,
g_{\text{ctrl}}^t,
g_{\text{cost}}^t),
\end{equation}
where $g_{\text{sec}}^t$ captures mitigation correctness, $g_{\text{lat}}^t$ captures latency and congestion impact, $g_{\text{ctrl}}^t$ captures controller-load pressure, and $g_{\text{cost}}^t$ captures operational cost. The scalar reward is:
\begin{equation}
R^t=\mathbf{w}^\top \mathbf{g}^t.
\end{equation}
Algorithm~\ref{alg:main} summarizes the complete fast-timescale control loop. The procedure \textsc{ReflectAndValidate} in Algorithm~\ref{alg:main} corresponds directly to Algorithm~\ref{alg:reflect}, which implements the slow-timescale LLM governance cycle, including evidence summarization, candidate $\Delta\Pi_k$ generation, stress-campaign validation, hard-safety checking, non-regression testing, and deployment gating.
\begin{algorithm}[H]
\caption{Two-Timescale Governance-Driven SDN-IoT Control}
\label{alg:main}
\footnotesize
\setlength{\baselineskip}{0.95\baselineskip}
\begin{algorithmic}[1]
\State Init switches/controller queues/flow capacities; init agents $\{\pi_{\theta_i},V_i\}$ and governance configuration $\Pi$
\State Set governance interval $K$ and rollout size $M$
\For{episode $=1\ldots E$}
  \State Reset dataplane queues/flow tables and controller backlog
  \For{timestep $t=0\ldots T$}
    \State Update controller: serve PacketIn jobs; apply pending FlowMod and meter-install jobs
    \For{each switch $i$}
      \State Observe $o_i^t$; sample $a_i^t\sim \pi_{\theta_i}(\cdot\mid o_i^t)$
      \State $a_i^{t*}\!\gets\!\textsc{SafetyFilter}(a_i^t,o_i^t,\Pi)$; execute; update $(q_i^{t+1},f_i^{t+1})$
      \State Update $(\rho_i^t,\varphi_i^t)$ and backlog $d^{t+1}$; compute $\mathbf{g}^t,R^t$
      \State Store $(o_i^t,a_i^t,a_i^{t*},R^t,o_i^{t+1})$
    \EndFor
    \If{buffer $\ge M$}
      \State PPO + critic update; clear buffer
    \EndIf
    \If{$t \bmod K = 0$}
      \State $\Pi \gets \textsc{ReflectAndValidate}(\Pi,\mathcal{T}_{t-K:t})$
    \EndIf
  \EndFor
\EndFor
\end{algorithmic}
\normalsize
\end{algorithm}
The reward components are tied to SDN-IoT operation. Specifically, $g_{\text{sec}}^t$ rewards correct mitigation and reduces missed attacks, $g_{\text{lat}}^t$ penalizes normalized queue growth and RTT inflation, $g_{\text{ctrl}}^t$ penalizes controller overload indicators, e.g., PacketIn drops, backlog, and FlowMod saturation, and $g_{\text{cost}}^t$ captures operational overhead, e.g., mirroring bandwidth and quarantine disruption.
Instead of optimizing only expected return, we adopt a risk-sensitive objective:
\begin{equation}
J(\pi)=
\max_{\pi}
\text{CVaR}_{\alpha}
\left(
\sum_{t=0}^{\infty}\gamma^tR^t
\right),
\end{equation}
which emphasizes worst-case trajectories and discourages catastrophic network states, e.g., persistent queue growth and control-plane collapse. In SDN-IoT, this is critical because rare overload episodes can dominate operator experience and cause large-scale disruption. The governance layer uses tail-risk evidence in $\mathcal{E}_k$ to increase controller-stress penalties and tighten the set of high-risk actions, thereby biasing learning and execution away from catastrophic outcomes.
Let $G \triangleq \sum_{t=0}^{\infty}\gamma^tR^t$ denote the discounted return. Conditional Value at Risk (CVaR) can be expressed via the Rockafellar-Uryasev~\cite{arletti2024directional} form:
\begin{equation}
\text{CVaR}_{\alpha}(G)
=
\min_{\eta\in\mathbb{R}}
\left[
\eta + \frac{1}{1-\alpha}\,\mathbb{E}\big[(G-\eta)_+\big]
\right].
\end{equation}
We clarify that CVaR is used in this work as a risk-sensitive design principle for reward shaping and policy governance, rather than as a direct CVaR-gradient estimator inside PPO. The PPO update remains the standard clipped surrogate objective described in Section \ref{Decentralized RL Control}. Risk sensitivity is operationalized by increasing the reward penalties associated with tail-risk indicators, including controller backlog peaks, PacketIn drops, RTT tail inflation, FlowMod bursts, and sustained queue growth. These quantities enter the reward vector through $g_{\text{ctrl}}^t$, $g_{\text{lat}}^t$, and $g_{\text{cost}}^t$, and their weights can be adjusted by the policy constitution $\Pi$. Thus, CVaR guides which failure modes are emphasized in shaping and governance, while PPO optimizes the resulting shaped reward using the standard update rule.
The LLM governance layer can tune $\mathbf{w}$ and related shaping parameters inside $\Pi$, for example, by increasing the weight of $g_{\text{ctrl}}^t$ after observing controller saturation. This provides a principled mechanism for shifting long-horizon priorities without modifying PPO parameters and destabilizing the RL optimization loop.

\subsection{Decentralized RL Control}
\label{Decentralized RL Control}
Each switch learns a local PPO policy $\pi_{\theta_i}(a_i|o_i)$ that maps switch-level telemetry to one SDN mitigation action.
A fast timestep $t$ corresponds to one telemetry-aggregation and actuation interval. During this interval, the controller processes pending PacketIn, FlowMod, and meter-install jobs; each switch observes $o_i^t$; the local PPO policy proposes $a_i^t$; the safety filter maps it to an executable action $a_i^{t*}$; and the environment advances to $s^{t+1}$. An episode is a finite closed-loop SDN-IoT run of length $T$ under a sampled traffic and attack scenario, producing the trajectory:
\begin{equation}
\mathcal{T}_{0:T}
=\{(o_i^t,a_i^t,a_i^{t*},R^t,o_i^{t+1})\}_{i=1,t=0}^{N,T-1}.
\end{equation}
The PPO agents are trained online within the SDN-IoT emulation loop using on-policy rollouts. Each agent collects rollout segments of length $M$ under the active governance configuration $\Pi_k$, computes advantage estimates, and updates its local actor and critic using the PPO objective. No offline replay buffer from earlier governance configurations is reused because such data correspond to a different effective environment after $\Pi$ changes.
Advantage estimation is computed as:
\begin{equation}
A_t=\sum_{k=0}^{\infty}(\gamma\lambda)^k
(r_{t+k}+\gamma V(o_{t+k+1})-V(o_{t+k})),
\end{equation}
and the clipped PPO surrogate objective is:
\begin{equation}
L^{\text{PPO}}=
\mathbb{E}\left[
\min(\rho_tA_t,\text{clip}(\rho_t,1-\epsilon,1+\epsilon)A_t)
\right],
\end{equation}
with:
\begin{equation}
\rho_t=
\frac{\pi_{\theta_i}(a_i^t|o_i^t)}
{\pi_{\theta_i}^{old}(a_i^t|o_i^t)}.
\end{equation}
Independent learning supports scalability, while shared reward terms and controller dynamics induce implicit coordination. Because $g_{\text{ctrl}}^t$ penalizes controller stress, each agent learns that excessive PacketIn, FlowMod churn, and meter-install activity can harm system-level stability.
The interaction between PPO learning and governance updates to $\Pi$ is handled through timescale separation. PPO inference and safety filtering occur at every fast timestep, PPO parameter updates occur after rollout segments of length $M$, and governance updates to $\Pi$ occur every $K$ timesteps, with $K \gg M$ in the evaluated setting. Thus, each PPO rollout is generated under a fixed governance configuration, while candidate $\Delta\Pi_k$ updates are generated and validated outside the fast control path. If accepted, $\Pi_{k+1}$ is activated only at a policy checkpoint; otherwise, the system continues with $\Pi_k$.
This design limits, but does not eliminate, non-stationarity. Changing $\Pi$ modifies the effective operating envelope through action masks, thresholds, reward shaping, and actuation caps. These changes are bounded, infrequent, and validated before deployment. PPO updates are computed only from rollouts collected under the currently active $\Pi_k$, preventing stale trajectories from older governance configurations from being mixed with the current update. Therefore, timescale separation should be interpreted as a practical stability-control mechanism rather than a formal convergence guarantee. Robustness outside the evaluated operating envelope depends on the diversity of stress campaigns, the boundedness of allowed governance edits, and the conservativeness of the acceptance gate.
The LLM governance layer does not replace PPO and intervene at every step. It modifies only the governance configuration that defines safe execution and reward-shaping parameters. This keeps PPO inference and safety filtering lightweight, while LLM governance runs infrequently and impacts only the compiled configuration $\Pi$ at checkpoints. Algorithm~\ref{alg:main} formalizes the fast-timescale loop, while the governance update stage is deferred to Section~\ref{subsec:llm_reflect} and Algorithm~\ref{alg:reflect}.

\subsection{Safety-Constrained Execution}
To ensure operational stability, PPO actions are filtered before execution:
\begin{equation}
a_i^{t*}=
\begin{cases}
a_i^t & a_i^t\in\mathcal{M}_{\Pi}(o_i^t),\\
\textsc{Fallback}(o_i^t,\Pi) & \text{otherwise}.
\end{cases}
\end{equation}
Here, $\mathcal{M}_{\Pi}(o_i^t)$ is the admissible action mask induced by $\Pi$, and $\textsc{Fallback}(o_i^t,\Pi)$ is a safe substitute selected when the PPO-proposed action violates controller-aware constraints.
The safety filter is not an independent learning policy. It is a deterministic execution guard that maps a proposed PPO action $a_i^t$ to an executable action $a_i^{t*}$ under the current governance configuration $\Pi$. This distinction is important because $\Pi$ does not replace the PPO policy $\pi_{\theta_i}$; instead, it defines the operating envelope within which PPO decisions are allowed to execute. For example, when PacketIn intensity $\rho_i^t$ and controller backlog $d^t$ are high, $\Pi$ may mask controller-heavy actions, e.g., \texttt{DROP\_FLOW}, \texttt{QUARANTINE}, and \texttt{RATE\_LIMIT}; when flow-table pressure $f_i^t$ is high, $\Pi$ may restrict actions that require additional FlowMod installation and increase rule churn.
If the governance layer produces an invalid delta, e.g., non-parsable output, schema violation, unresolved constraint conflict, and a candidate configuration failing hard-safety checks, then $\Pi$ is not updated, and the system continues with $\Pi_k$. Likewise, the safety filter always maps infeasible PPO actions to a valid fallback action, ensuring that malformed LLM outputs and unsafe candidate updates cannot cause undefined behavior in the dataplane and controller. The safest action is context-dependent: under severe controller overload, avoiding controller-intensive rule changes may be safest, whereas under low load, stronger containment can be admissible.
The admissible action mask is:
\begin{equation}
\mathcal{M}_{\Pi}(o)
=
\{a\in\mathcal{A}\;:\;\mathbf{c}(o,a;\boldsymbol{\tau},\boldsymbol{\chi})\le \mathbf{0}\},
\end{equation}
where $\mathbf{c}$ encodes controller-aware execution constraints, including backlog caps, FlowMod throttles, meter-install limits, flow-table pressure limits, and action-specific actuation caps. Thus, updates to $(\boldsymbol{\tau},\boldsymbol{\chi})$ directly reshape the feasible action set without changing PPO parameters.
In the discrete action space, the safety filter selects the closest feasible action under a task-specific substitution cost $D$:
\begin{equation}
a_i^{t*}
=
\arg\min_{a\in\mathcal{M}_{\Pi}(o_i^t)} D(a,a_i^t).
\end{equation}
Here, $D(a,a_i^t)$ measures the operational difference between the original PPO action and a feasible substitute. For example, replacing \texttt{QUARANTINE} with \texttt{RATE\_LIMIT} may be considered closer than replacing it with \texttt{ALLOW}, because both actions still mitigate suspicious traffic but differ in controller cost and disruption severity. This formulation clarifies that the safety filter reduces unsafe actuation while preserving the intent of the PPO action as much as possible.
Algorithm~\ref{alg:reflect} specifies the slow-timescale implementation of the \textsc{ReflectAndValidate} procedure called in Algorithm~\ref{alg:main}. It receives the current governance configuration $\Pi_k$ and recent trace window $\mathcal{T}_k$, generates a candidate policy delta $\Delta\Pi_k$, evaluates it under targeted stress campaigns, and deploys it only if hard-safety and non-regression checks are satisfied.
\begin{algorithm}[H]
\caption{ReflectAndValidate: LLM Governance Update of $\Pi$}
\label{alg:reflect}
\footnotesize
\begin{algorithmic}[1]
\Require Current governance configuration $\Pi_k$, recent trace window $\mathcal{T}_k$
\Ensure Updated governance configuration $\Pi_{k+1}$

\Statex \textbf{// 0) Evidence bundle}
\State $\mathcal{E}_k \gets \textsc{SummarizeTrace}(\mathcal{T}_k)$
\Comment{QoS tails, PacketIn drops, backlog, FlowMods, churn, FP/FN, operating conditions}

\Statex \textbf{// 1) Diagnose and compile bounded governance edits}
\State $\mathcal{I}_k \gets \textsc{CriticLLM}(\Pi_k,\mathcal{E}_k)$
\State $\Delta\Pi_k^{raw} \gets \mathcal{I}_k.\texttt{proposals}$
\State $\Delta\Pi_k \gets \textsc{CompilerLLM}(\Delta\Pi_k^{raw})$
\Comment{masks, thresholds, reward patches, caps}

\Statex \textbf{// 2) Normalize, validate, and merge}
\State $\Delta\Pi_k \gets \textsc{Normalize}(\Delta\Pi_k)$
\State $\Delta\Pi_k \gets \textsc{Deduplicate}(\Delta\Pi_k)$
\State $\Delta\Pi_k \gets \textsc{SchemaCheck}(\Delta\Pi_k)$
\State $\Pi_{cand} \gets \Pi_k \oplus \Delta\Pi_k$
\Comment{hard constraints $\rightarrow$ masks $\rightarrow$ thresholds $\rightarrow$ reward patches}

\Statex \textbf{// 3) Stress-test current and candidate configurations}
\State $\mathcal{C}_{rt} \gets \textsc{RedTeamLLM}(\mathcal{E}_k)$
\State $\mathcal{M}_k \gets \textsc{EvaluatePolicy}(\Pi_k,\mathcal{C}_{rt})$
\State $\mathcal{M}_{cand} \gets \textsc{EvaluatePolicy}(\Pi_{cand},\mathcal{C}_{rt})$
\State $\Delta\mathcal{M} \gets \mathcal{M}_{cand}-\mathcal{M}_k$

\Statex \textbf{// 4) Gate deployment}
\State $\texttt{safe} \gets \textsc{HardSafety}(\Pi_{cand})$
\State $\texttt{ok} \gets \textsc{NonRegression}(\Delta\mathcal{M})$
\State $\texttt{approve} \gets \textsc{JudgeLLM}(\Pi_k,\Pi_{cand},\Delta\mathcal{M},\texttt{safe},\texttt{ok})$

\If{$\texttt{approve} \wedge \texttt{safe} \wedge \texttt{ok}$}
  \State $\Pi_{k+1} \gets \Pi_{cand}$
\Else
  \State $\Pi_{k+1} \gets \Pi_k$
\EndIf

\Statex \textbf{// 5) Reproducibility logging}
\State $\textsc{LogArtifacts}(\Pi_k,\Delta\Pi_k,\Pi_{cand},\mathcal{C}_{rt},\mathcal{M}_k,\mathcal{M}_{cand},\Pi_{k+1})$
\State \Return $\Pi_{k+1}$
\end{algorithmic}
\normalsize
\end{algorithm}

\subsection{LLM-Based Governance Policy Evolution}
\label{subsec:llm_reflect}
We maintain a structured governance constitution $\Pi$ containing admissible action masks, controller-aware thresholds, reward priorities, safety rules, and actuation limits. Given a trajectory window $\mathcal{T}_k$, the governance update operator produces:
\begin{equation}
\Pi_{k+1}=\mathcal{R}(\Pi_k,\mathcal{T}_k),
\end{equation}
with deployment governed by:
\begin{equation}
\Pi^*=
\begin{cases}
\Pi_{k+1}, & \text{if validated},\\
\Pi_k, & \text{otherwise}.
\end{cases}
\end{equation}
The LLM governance layer operates on the slow timescale and revises the explicit policy constitution $\Pi$ rather than controlling packet-level actions. It receives summarized trajectory data, including controller backlog, PacketIn drops, FlowMod intensity, RTT tails, false positives and false negatives, and action-churn indicators. It does not access raw packet payloads and modify PPO parameters. Its role is to diagnose recurring failure modes, propose bounded policy edits, and validate those edits before deployment.
The governance layer consists of five functional roles: Critic, Compiler, Red-Team, Judge, and Memory. The Critic identifies failure modes; the Compiler converts recommendations into policy deltas; the Red Team generates targeted stress campaigns; the Judge approves or rejects candidate updates; and the Memory component stores accepted updates, rejected candidates, validation outcomes, and historical evidence for auditability.
To formalize governance edits, we define:
\begin{equation}
\Delta\Pi_k \triangleq
\big(
\Delta\mathcal{S}_{\Pi},
\Delta\boldsymbol{\tau},
\Delta\mathbf{w},
\Delta\boldsymbol{\chi}
\big),
\end{equation}
where $\Delta\mathcal{S}_{\Pi}$ updates admissible action masks, $\Delta\boldsymbol{\tau}$ updates controller-aware thresholds, $\Delta\mathbf{w}$ updates reward priorities, and $\Delta\boldsymbol{\chi}$ updates auxiliary policy patches, e.g., action-dependent costs, throttles, and actuation caps.
The candidate policy is obtained through:
\begin{equation}
\Pi_{cand}=\Pi_k \oplus \Delta\Pi_k,
\end{equation}
where $\oplus$ applies precedence rules, conflict resolution, deduplication, and range clipping. Hard constraints take precedence over reward patches, ensuring conservative policy evolution.
For example, if frequent \texttt{DROP\_FLOW} and \texttt{QUARANTINE} actions increase controller backlog and delay rule activation, the governance layer may propose a $\Delta\Pi_k$ that masks controller-heavy actions during overload, increases controller-stress penalties, and validates the update under burst, mimicry, and multi-switch stress campaigns. If validation fails, the system retains $\Pi_k$.
Let $\mathcal{M}(\Pi,\mathcal{C})$ return a metric vector:
\begin{equation}
\mathcal{M}=
(F1,\ \text{RTT}_{p95},\ D_{ctrl},\ \text{FlowMods},\ \text{DropRate}).
\end{equation}
Candidate policies are accepted only if they satisfy hard-safety and non-regression requirements:
\begin{align}
\textsc{HardSafety}(\Pi_{cand}) &= 1,\\
\textsc{NonRegression}(\Delta\mathcal{M})
&=
\mathbb{I}
\Big[
\Delta F1 \ge -\epsilon_{F1}
\nonumber\\
&\qquad\wedge\;
\Delta \text{RTT}_{p95} \le \epsilon_{rtt}
\nonumber\\
&\qquad\wedge\;
\Delta D_{ctrl} \le \epsilon_{ctrl}
\Big].
\end{align}
where
\begin{equation}
\Delta\mathcal{M}
=
\mathcal{M}(\Pi_{cand},\mathcal{C}_{rt})
-
\mathcal{M}(\Pi_k,\mathcal{C}_{rt}).
\end{equation}
The deployment rule becomes:
\begin{equation}
\Pi_{k+1}=
\begin{cases}
\Pi_{cand},
&
\texttt{safe}
\wedge
\texttt{nonreg},
\\
\Pi_k,
&
\text{otherwise}.
\end{cases}
\end{equation}
Importantly, only the candidate policy-delta generation is LLM-dependent. Parsing, schema validation, normalization, deduplication, range clipping, merge precedence, conflict resolution, hard-safety checking, and non-regression testing are performed by deterministic code. Consequently, non-determinism is confined to candidate generation, whereas deployment decisions remain deterministic and fail-closed. If any validation stage fails, the previously validated configuration $\Pi_k$ is retained.
Implementation-level safeguards further improve auditability. Each governance role uses a fixed system prompt and a typed output contract. In the reference implementation, the Critic, Compiler, Red-Team, and Judge roles use the same backbone model identifier, \texttt{gpt-4o-mini}, with fixed decoding parameters. All prompts, responses, hashes, latency values, parse outcomes, and accepted and rejected deltas are logged for reproducibility.
The validation mechanism is empirical rather than formally verified. Hard safety and non-regression checks reduce the likelihood of unsafe updates in the evaluated stress campaigns, but they do not guarantee safety under different conditions. Therefore, accepted $\Delta\Pi_k$ updates should be interpreted as conservative empirical refinements. The acceptance gate also acts as an anti-oscillation mechanism by allowing deployment only when candidate updates satisfy conservative validation criteria.

\section{Threat Model}
\label{sec:threat}
We consider adversarial SDN-IoT networks in which malicious traffic aims to degrade network stability, evade mitigation, and overload the control plane. The defender is the proposed multi-agent control solution, consisting of decentralized RL switch agents, a centralized SDN controller, and an LLM governance layer that evolves the global policy constitution $\Pi$.
We model a traffic-only adversary. The attacker interacts with the system through packet and flow generation patterns, but does not compromise SDN control software, switch firmware, telemetry counters, communication channels, the training pipeline, PPO parameters, reward signals, and the governance layer. This threat model targets the dominant operational risk in SDN-IoT: traffic-induced control-plane and queueing collapse. Platform-compromise attacks are treated as out of scope.
Let $s^t \in \mathcal{S}$ denote the latent network state and let $\mathbf{z}^t$ denote traffic, including benign IoT activity and adversarial behavior. Since the attacker acts only through traffic, the system evolves according to:
\begin{equation}
s^{t+1} \sim P(\cdot \mid s^t,\mathbf{a}^t,\mathbf{z}^t),
\end{equation}
where $\mathbf{a}^t$ is the joint mitigation action chosen by the switch agents. Thus, attacks indirectly affect future observations by changing queue lengths, controller workload, flow-table pressure, and rule-installation delays.
The attacker may adapt $\mathbf{z}^t$ online based on observable network responses, e.g., RTT, packet loss, throughput, and reachability. However, it cannot directly observe the internal controller backlog, reward decomposition, PPO parameters, and the current $\Pi$ parameters, except indirectly through data-plane behavior. The adversarial objective is:
\begin{equation}
\max_{\mathbf{z}^{0:T}}
\sum_{t=0}^{T-1}
\Big(
\lambda_1 D(s^t,\mathbf{a}^t)
-
\lambda_2 \mathbb{I}[\text{detected}]
\Big),
\end{equation}
where $D(\cdot)$ represents disruption measures, e.g., queue growth, RTT inflation, and controller saturation. Detection provides operational visibility through defender-triggered containment, e.g., sustained \texttt{DROP\_FLOW}, \texttt{QUARANTINE}, and persistent \texttt{RATE\_LIMIT}, thereby reducing attacker throughput and increasing attacker cost.
The attacker can generate distributed, bursty, temporally correlated, and mimicry-based traffic. These patterns indirectly impact key system variables, including switch queues $q_i^t$, PacketIn rates $\rho_i^t$, flow-table pressure $f_i^t$, and controller backlog $d^t$. We explicitly evaluate five SDN-specific stress patterns: 
\begin{enumerate}
\item control-plane saturation attacks that induce flow misses and churn;
\item low-and-slow persistent attacks that remain below static thresholds while accumulating tail risk;
\item burst-and-idle campaigns that exploit delayed actuation and queue dynamics;
\item mimicry attacks that imitate legitimate synchronized IoT bursts, e.g., firmware updates and telemetry waves;
\item multi-switch correlated campaigns that amplify shared-controller bottlenecks.
\end{enumerate}
Switch agents observe only local telemetry:
\begin{equation}
o_i^t \sim O_i(\cdot \mid s^t).
\end{equation}
Thus, the defender operates under partial observability, while the attacker has only black-box access to data-plane effects. This asymmetry is realistic in SDN-IoT deployments, where attackers can probe network behavior but cannot access the controller's internals.
The SDN controller, switch software, telemetry counters, and authenticated controller-switch channels are assumed trusted. The attacker cannot inject controller rules, tamper with counters, poison evaluation labels, alter reward signals, and interfere with PPO optimization. The governance layer consumes deterministic aggregate counters and validated trace summaries rather than raw packet payloads, reducing exposure to prompt manipulation. Nevertheless, adversarial traffic can still skew aggregate evidence statistics and may indirectly affect candidate $\Delta\Pi$ updates via distribution shift.
Out-of-scope threats include controller compromise, switch firmware compromise, telemetry tampering, counter spoofing, training-time poisoning, evaluation-label poisoning, cryptographic attacks on authenticated channels, and direct compromise of the external LLM provider. These deployment risks are important but orthogonal to the traffic-induced closed-loop instability targeted in this work.
If the LLM service is unavailable, produces invalid outputs, or proposes unsafe updates, the system fails closed by retaining the previously validated governance configuration $\Pi_k$. In addition, the LLM governance layer receives only summarized telemetry and trace statistics, not raw packet payloads and privileged controller secrets. These design choices reduce, but do not eliminate, the attack surface introduced by external LLM services. Attacks that compromise the external LLM provider, leak telemetry through the API, and manipulate the service infrastructure are considered deployment risks and are discussed as limitations rather than fully solved threats in this work.
The defender aims to minimize long-horizon tail risk:
\begin{equation}
\min_{\pi,\Pi}
\text{CVaR}_{\alpha}
\left(
\sum_{t=0}^{T-1}\gamma^t R^t
\right).
\end{equation}
This objective prioritizes rare but catastrophic outcomes, e.g., persistent congestion, controller overload, queue explosions, high-percentile RTT inflation, and widespread misclassification. When tail evidence appears, the LLM governance layer may reweight controller-stress penalties and tighten feasibility constraints for controller-heavy actions, thereby prioritizing resilience under adaptive traffic.
Accordingly, CVaR-related behavior is evaluated ex-post using tail metrics, e.g., worst-case agent degradation, controller-backlog peaks, RTT$_{p95}$ inflation, and catastrophic overload frequency, rather than through a separate CVaR-gradient optimization procedure.

\section{Experimental Evaluation Protocol}
\label{sec:experimental_setup}
\begin{figure}[t] 
\centering
\includegraphics[width=0.50\textwidth]{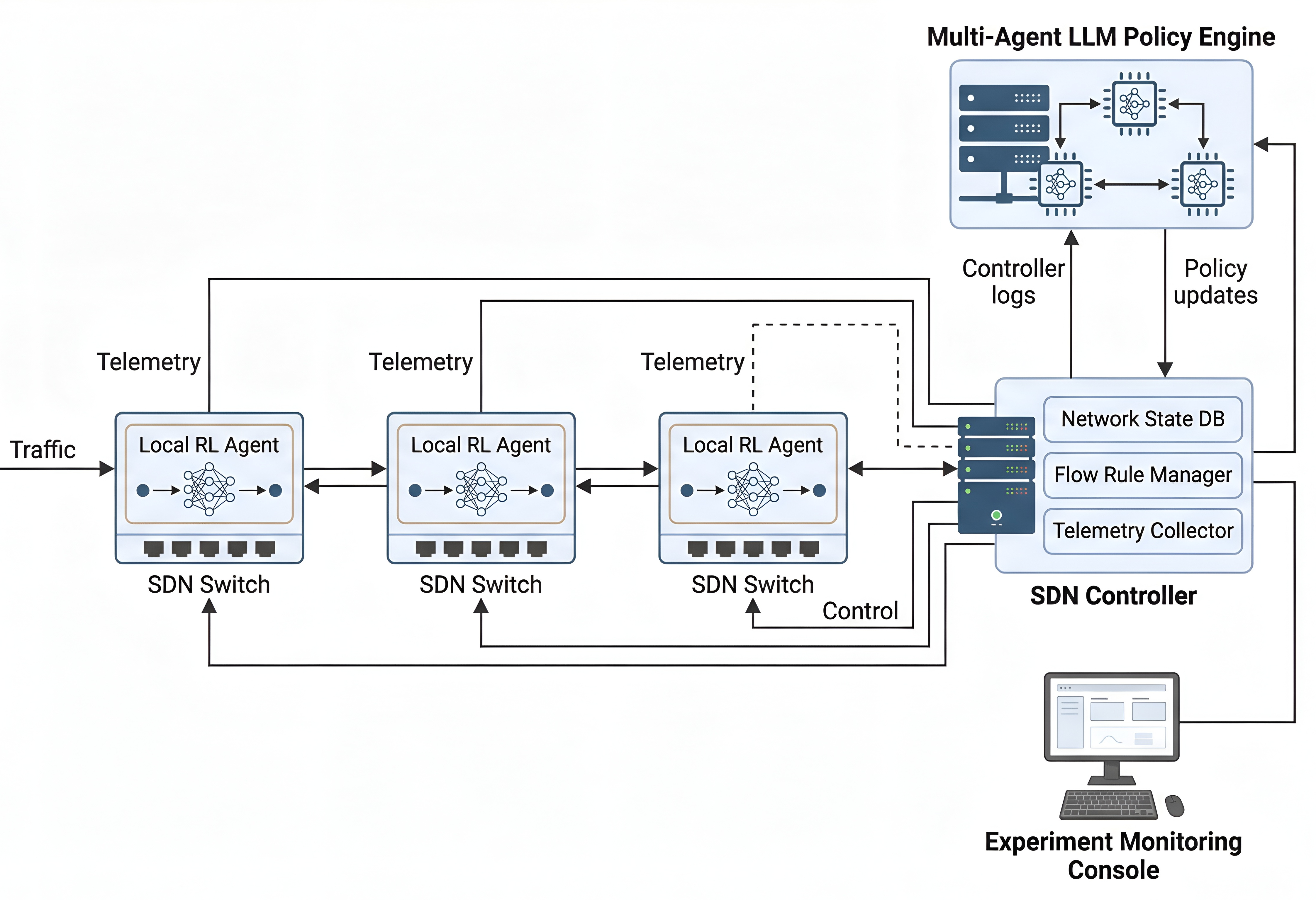} 
\caption{Experimental architecture of the two-timescale governance-driven SDN-IoT network. Local PPO agents run on SDN switches; the controller aggregates telemetry and enforces flow rules; and the LLM governance layer performs structured policy evolution based on controller logs.}
\label{fig:system_architecture}
\end{figure}

Experimental validation uses the integrated SDN-IoT architecture in Figure~\ref{fig:system_architecture}. IoT traffic is processed by decentralized PPO agents on SDN switches under the control of a centralized controller, while the LLM governance layer updates $\Pi$. The evaluation is framed as a closed-loop systems study that models controller saturation, delayed rule actuation, PacketIn buffering, FlowMod and meter-install activity, and flow-table churn. The experiments were conducted in a controlled SDN-IoT emulation testbed on a single Windows-based workstation/server using WSL2 for Linux-based SDN emulation. The same host executed the SDN controller, virtual switches, IoT, adversarial traffic generators, PPO agents, safety filter, telemetry collector, governance-validation pipeline, and logging modules. This single-server configuration was used to maintain consistent resource allocation and isolate closed-loop control behavior from inter-server communication variability. The topology used one centralized SDN controller connected to $N$ OpenFlow-enabled edge switches, with $M$ IoT devices attached to each switch. Thus, the emulated network contained $N \times M$ IoT hosts. Adversarial hosts were distributed across multiple switch domains to evaluate localized, distributed, and multi-switch correlated attacks. The topology was selected to expose shared controller backlog, PacketIn amplification, delayed FlowMod activation, flow-table pressure, queue accumulation, and cross-switch attack coupling.
\begin{table}[t]
\centering
\scriptsize
\caption{Experimental testbed and topology configuration.}
\label{tab:experimental_testbed}
\begin{tabular}{ll}
\toprule
\textbf{Component} & \textbf{Configuration} \\
\midrule
Host platform & Windows workstation with WSL2/Ubuntu \\
Physical servers & 1 dedicated workstation/server \\
SDN emulator & Mininet/Containernet-based OpenFlow emulation \\
SDN controller & Ryu/ONOS/Floodlight controller \\
Controller instances & 1 centralized controller \\
SDN switches & $N$ OpenFlow-enabled edge switches \\
IoT devices & $M$ IoT hosts per switch; $N \times M$ total hosts \\
Topology & Multi-switch SDN-IoT access topology \\
Traffic types & Benign telemetry, synchronized bursts, adversarial flows \\
Attack scenarios & Saturation, low-and-slow, burst-idle, mimicry, correlated attacks \\
RL implementation & Python/PyTorch PPO actor--critic agents \\
Governance interval & Every $K$ fast timesteps \\
Episode horizon & $T$ fast timesteps \\
\bottomrule
\end{tabular}
\end{table}
A fast timestep corresponds to one telemetry-aggregation and actuation interval. All methods use the same timestep duration, telemetry interval, controller service-rate configuration, queue capacity, and episode horizon, making Figures~\ref{fig:rl_fast}--\ref{fig:rl_stable} directly comparable. An episode is a finite, closed-loop run of length $T$ under a single sampled benign-traffic-and-attack scenario. A trajectory window $\mathcal{T}_{t-K:t}$ is the latest $K$-step segment used by the LLM governance layer to summarize evidence and propose candidate $\Delta\Pi_k$ updates.
Traffic is generated inside the emulation environment using controlled benign IoT models and parameterized adversarial injection processes. Static public intrusion datasets are not used because they cannot capture mitigation--controller feedback loops central to closed-loop stability analysis~\cite{ficco2017security,zhu2023secure,xing2024security,huang2024security}. The controller is modeled with bounded service rate $\mu$, stochastic service times, bounded PacketIn buffering, delayed FlowMod execution, meter-install delays, and finite flow-table capacity.
Benign traffic includes periodic sensing, background IoT communication, and synchronized bursts, while adversarial traffic includes control-plane saturation, low-and-slow, burst-and-idle, mimicry, and multi-switch correlated attacks.
PPO agents are trained online using on-policy rollout segments collected under the active governance configuration $\Pi_k$. PPO actor--critic updates use only trajectories generated under that same configuration; stale trajectories from older governance configurations are not reused after $\Pi$ changes.
All compared methods use identical topologies, traffic seeds, controller settings, observation spaces, and action sets. Baselines include static threshold-based mitigation, unconstrained PPO, and constrained PPO without governance. Performance is evaluated across security, QoS, and operational cost using macro-F1, worst-case agent performance, RTT percentiles, FlowMod intensity, disruption, and controller backlog evolution.
Disruption is defined as the fraction of benign traffic impacted by mitigation actions:
\begin{equation}
\text{Disruption}
=
\frac{
N_{\text{benign, impacted}}
}{
N_{\text{benign}}
}
\end{equation}
where $N_{\text{benign}}$ is the total number of benign flows and sessions, and $N_{\text{benign, impacted}}$ is the subset impacted by rate limiting, dropping, quarantine, and equivalent service degradation. Lower disruption indicates lower collateral impact.
Multiple independent seeds are used, and comparisons are conducted using paired statistical tests with effect sizes. Governance is evaluated every $K$ fast timesteps. At each checkpoint, the LLM governance layer receives $\mathcal{T}_{t-K:t}$, generates candidate $\Delta\Pi_k$, and validates it against fixed red-team campaigns with matched seeds for both $\Pi_k$ and $\Pi_k\oplus\Delta\Pi_k$. Candidate updates are accepted only after schema validity, hard safety, and conservative non-regression checks.
For transparency, the evaluation reports proposed governance updates, accepted updates, rejected valid updates, and invalid deltas. It also summarizes changed components of $\Pi$, including action-mask, threshold, reward-weight, and actuation-cap changes. Invalid deltas include parsing failures, schema violations, and unresolved conflicts, whereas valid deltas that are rejected fail hard-safety and non-regression criteria.
All governance rounds log evidence bundles, parsed deltas, validation metrics, approval decisions, invalid-output counts, rejected-delta counts, latency, and overhead to support auditability and replayability.

\subsection{Adversarial Scenario Design}
\label{subsec:adversarial_design}
Adversarial scenarios are designed to stress both the data plane and the control plane by exercising controller backlog, delayed rule installation, and flow-table churn. Attack patterns are injected into heterogeneous IoT traffic and vary across episodes in intensity, synchronization probability, and temporal structure.
We consider three primary attack classes: high-volume burst attacks that trigger PacketIn surges and rule-install delays; distributed low-rate scans that increase flow-table churn without exceeding volume thresholds; and synchronized mimicry patterns that resemble benign IoT update bursts while gradually elevating controller backlog.
For interpretability, each scenario is associated with a dominant stress mechanism. High-volume burst attacks primarily stress queue growth and PacketIn buffering. Distributed low-rate scans primarily stress flow-table churn and FlowMod generation. Synchronized mimicry patterns test whether the policy avoids false containment of benign-looking IoT bursts. Burst-and-idle attacks test backlog hysteresis and delayed actuation, while multi-switch correlated attacks test shared-controller saturation. Therefore, the reported metrics should be interpreted relative to the controller, queueing, and flow-table mechanism stressed by each scenario.
Attack intensity is parameterized through injection multipliers and device synchronization probability, ranging from mild perturbation with controller utilization below $60\%$ to near-saturation conditions where $\sum_i \rho_i^t \rightarrow \mu$. Attack timing is randomized to reduce temporal overfitting. Validation campaigns further include burst-and-idle attacks, multi-switch correlated attacks, and low-and-slow persistent patterns that accumulate tail risk without immediate alarms.
Several scenarios are configured near stability boundaries to test whether policies maintain bounded backlog and QoS under escalation. Infrastructure parameters, including controller service rate, queue size, and flow-table capacity, remain fixed across stress levels so that degradation is attributable to traffic characteristics rather than topology changes. Benign synchronized bursts are included in both training and evaluation to penalize unnecessary disruption of legitimate IoT behavior.

\subsection{Security Stabilization Dynamics}
\label{subsec:stabilization}
\begin{figure}[t]
    \centering
    \includegraphics[width=\columnwidth]{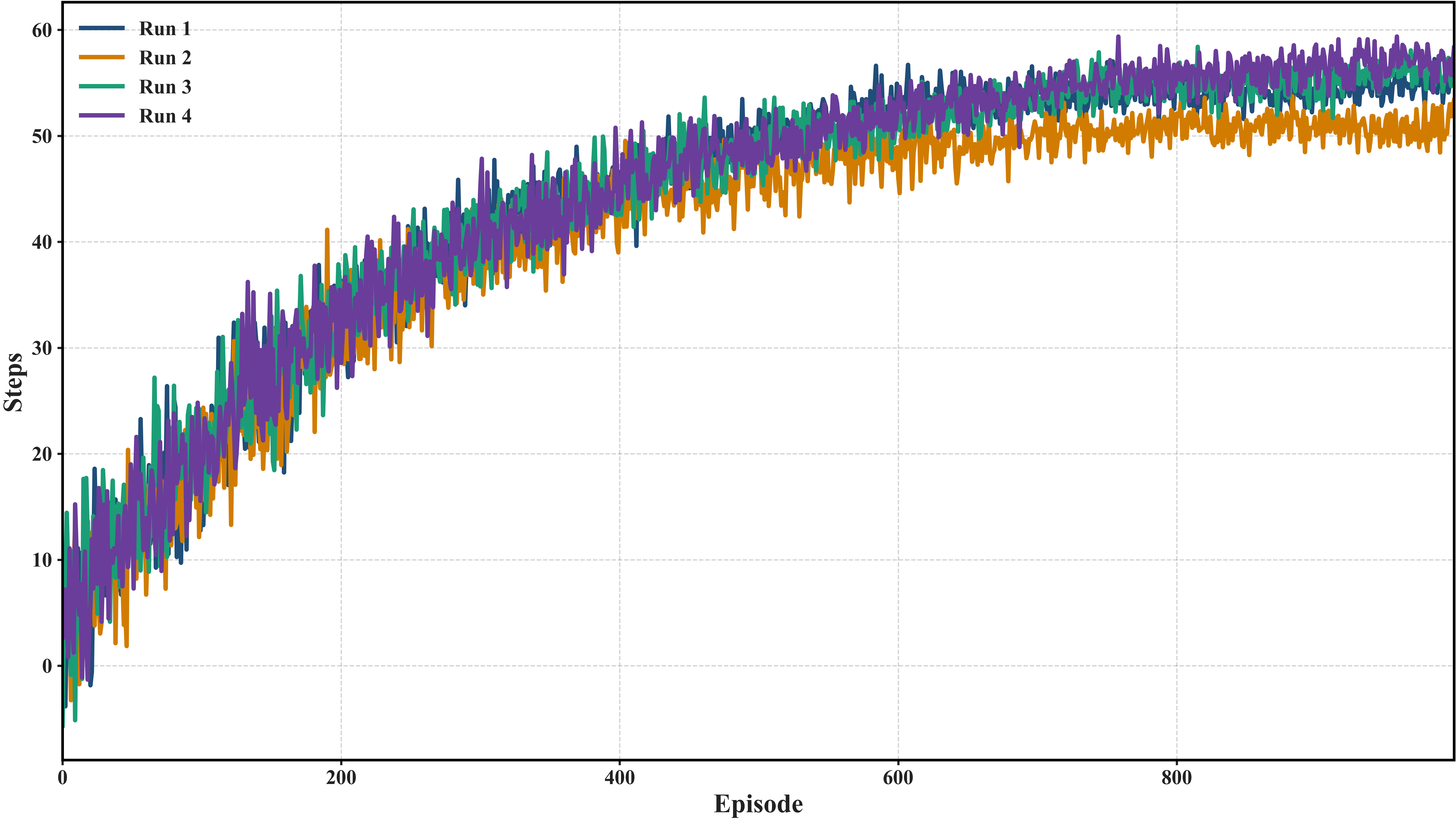}
    \caption{Security stabilization dynamics under the Fast configuration (PPO clip $\epsilon=0.30$). Steps per episode indicate sustained adversarial containment without control-plane overload.}
    \label{fig:rl_fast}
\end{figure}

\begin{figure}[t]
    \centering
    \includegraphics[width=\columnwidth]{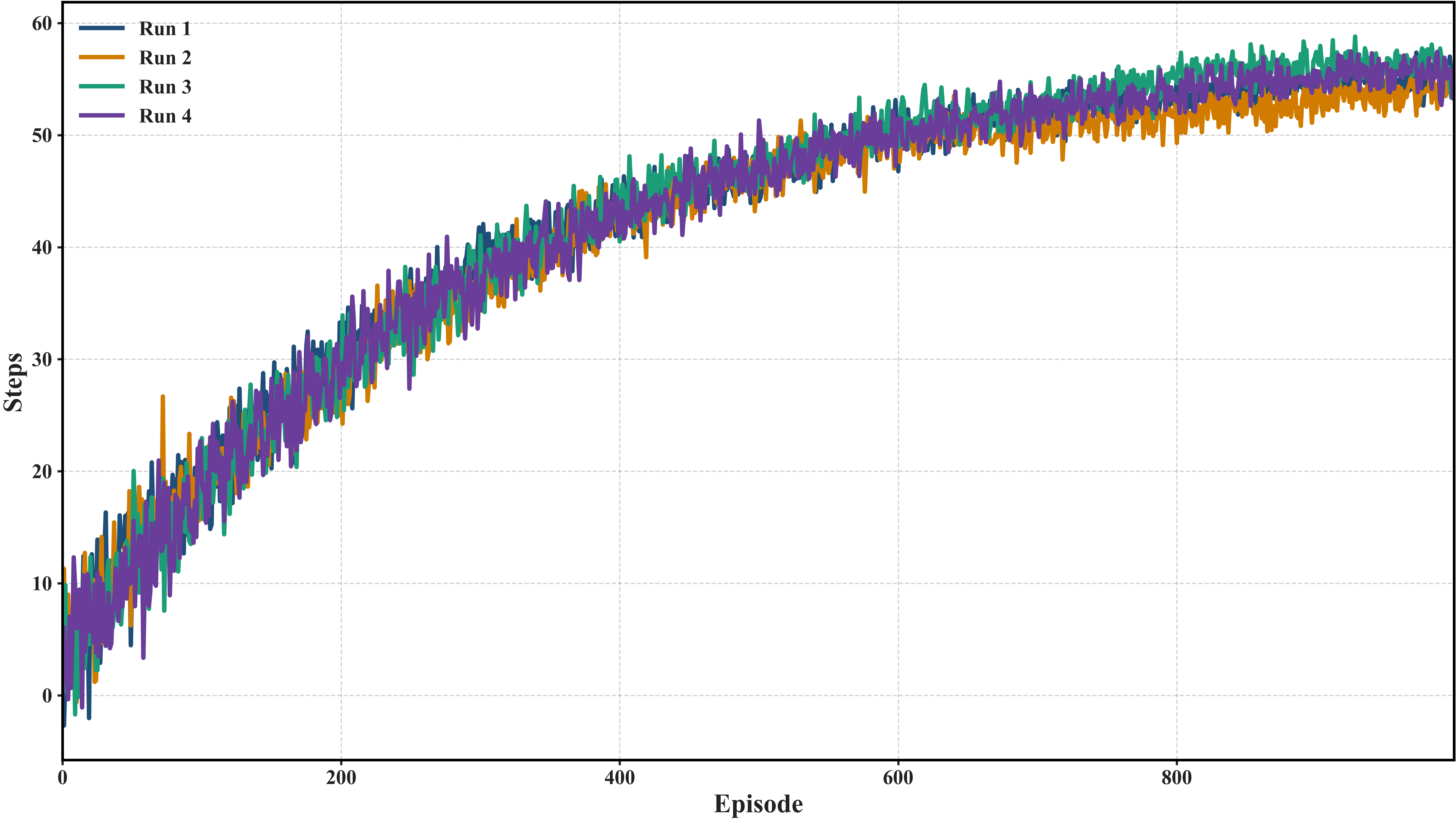}
    \caption{Security stabilization dynamics under the Best Performance configuration (PPO clip $\epsilon=0.20$). The highest asymptotic episode length reflects extended safe containment under adversarial traffic while preserving controller stability.}
    \label{fig:rl_best}
\end{figure}

\begin{figure}[t]
    \centering
    \includegraphics[width=\columnwidth]{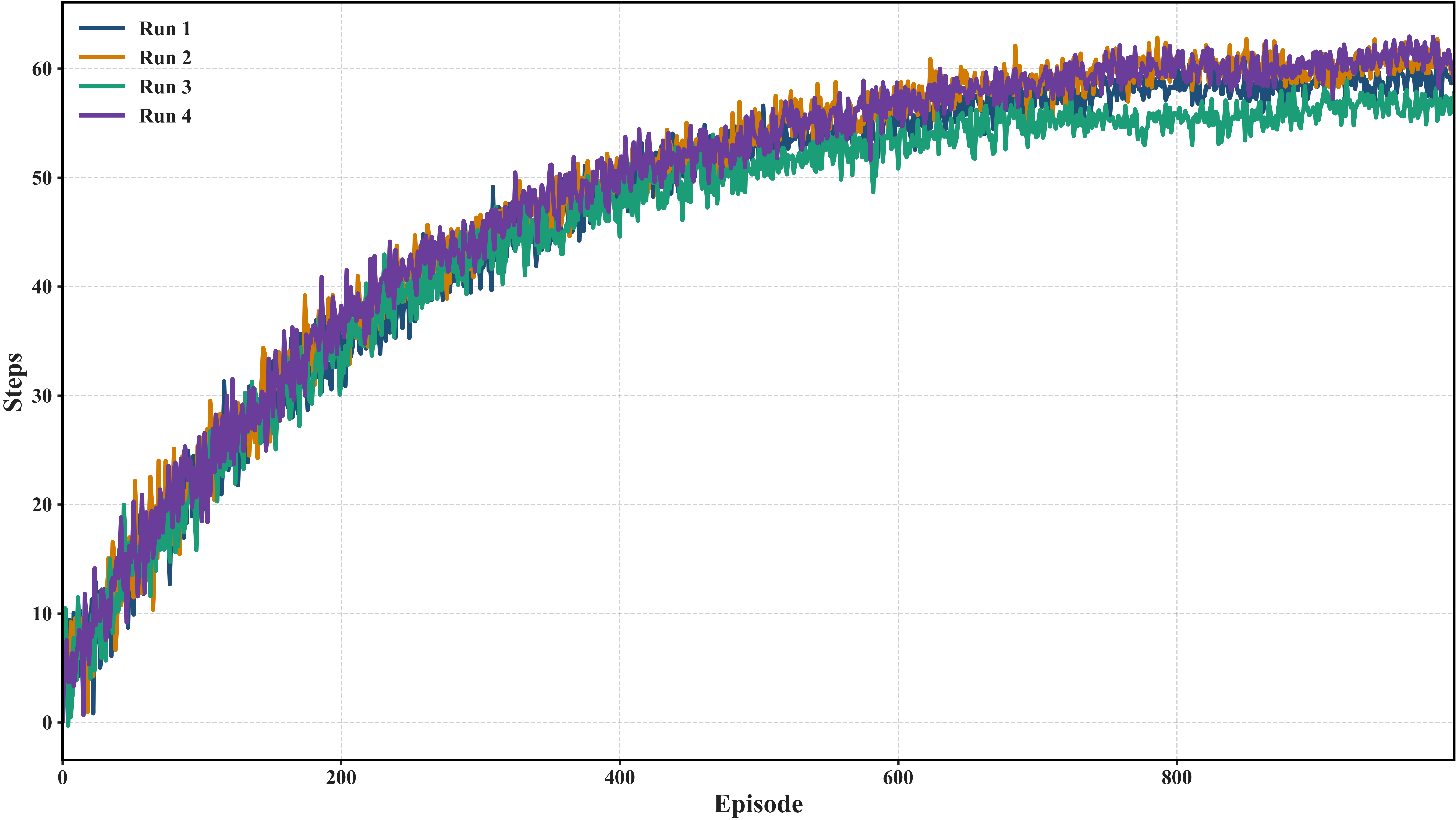}
    \caption{Security stabilization dynamics under the Stable configuration (PPO clip $\epsilon=0.10$). Reduced cross-run variance indicates robust mitigation behavior under stochastic traffic with bounded control-plane impact.}
    \label{fig:rl_stable}
\end{figure}

Figures~\ref{fig:rl_fast}--\ref{fig:rl_stable} show the training dynamics of decentralized PPO mitigation agents under three clipping settings: Fast ($\epsilon=0.30$), Best Performance ($\epsilon=0.20$), and Stable ($\epsilon=0.10$). The PPO clipping parameter controls the trust-region width by constraining the likelihood ratio $\rho_t$ to $[1-\epsilon,1+\epsilon]$. Larger $\epsilon$ enables faster adaptation but higher variance, whereas smaller $\epsilon$ yields more conservative and stable updates.
All configurations are evaluated using the same fixed-duration timestep, telemetry aggregation interval, controller service-rate configuration, queue capacity, and episode horizon. Therefore, the episode-length curves in Figures~\ref{fig:rl_fast}--\ref{fig:rl_stable} are directly comparable. The horizontal axis represents training episodes, each containing the same maximum number of observe--decide--filter--actuate--update intervals. Episode length is interpreted as a long-horizon stability proxy: longer episodes indicate that the system sustains safe operation for more control intervals before termination due to controller overload, persistent backlog growth, and severe QoS degradation.
Because mitigation actions impact controller backlog, flow-table churn, and delayed rule activation, abrupt policy shifts can amplify control-plane instability. Across all $\epsilon$ settings, episode length improves smoothly and reaches a plateau after approximately 700 episodes, indicating bounded closed-loop behavior without divergence.
The plateau should not be interpreted as an artifact of unequal episode duration because all methods use the same timestep duration and episode horizon. Instead, it reflects convergence toward a controller-capacity-limited operating regime. Early training suppresses high-impact malicious flows and obvious PacketIn escalation, while later training faces harder low-rate and mimicry patterns as PacketIn and FlowMod pressure approach the controller service envelope. The best-performing configuration reaches a higher and lower-variance plateau, indicating that controller-aware reward shaping and $\Pi$-based safety filtering prevent convergence to a higher-backlog equilibrium.
The Fast configuration stabilizes earliest but shows higher cross-seed variance due to larger policy updates. The Best Performance configuration achieves the highest asymptotic episode length, balancing adaptation and regularization. The Stable configuration converges more gradually with lower variance, reflecting stronger trust-region regularization under backlog stress. Overall, bounded convergence across all settings shows that clipped PPO, combined with $\Pi$-based safety filtering and controller-aware reward shaping, provides a stable RL foundation for two-timescale governance.

\section{System-Level Performance Evaluation}
\label{System-Level Performance Evaluation}
This section evaluates the proposed solution as a closed-loop SDN-IoT control system, focusing on security robustness, QoS stability, overhead, tail risk, and governance convergence.

\subsection{Detection Performance and Multi-Agent Consistency}
\label{subsec:results_detection}
\begin{figure}[t]
  \centering
  \includegraphics[width=0.80\columnwidth]{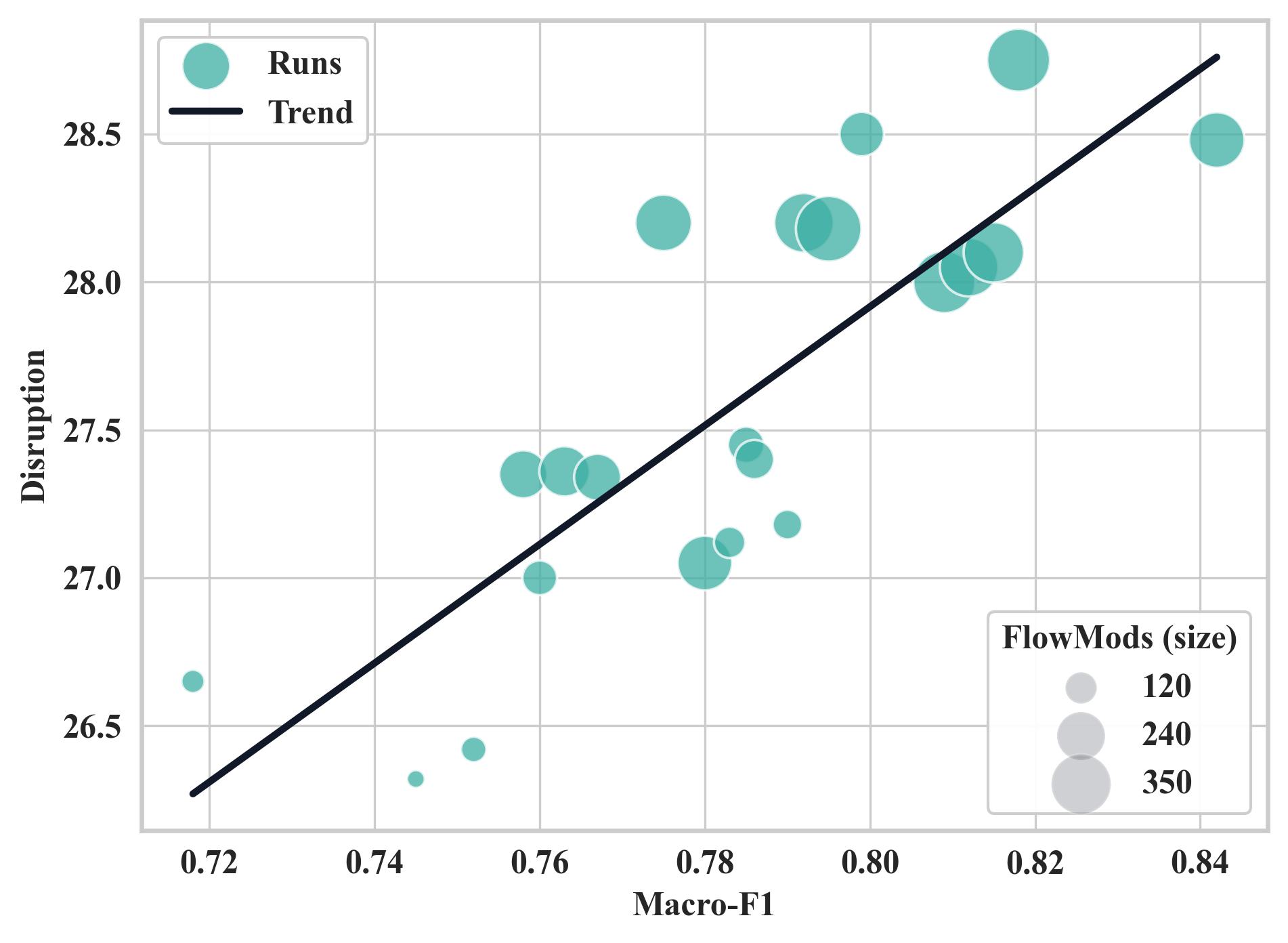}
  \caption{Empirical cumulative distribution function (CDF) of detection performance (Macro-F1) across evaluation episodes.}
  \label{fig:cdf_detection_f1}
\end{figure}

\begin{figure}[t]
  \centering
  \includegraphics[width=0.80\columnwidth]{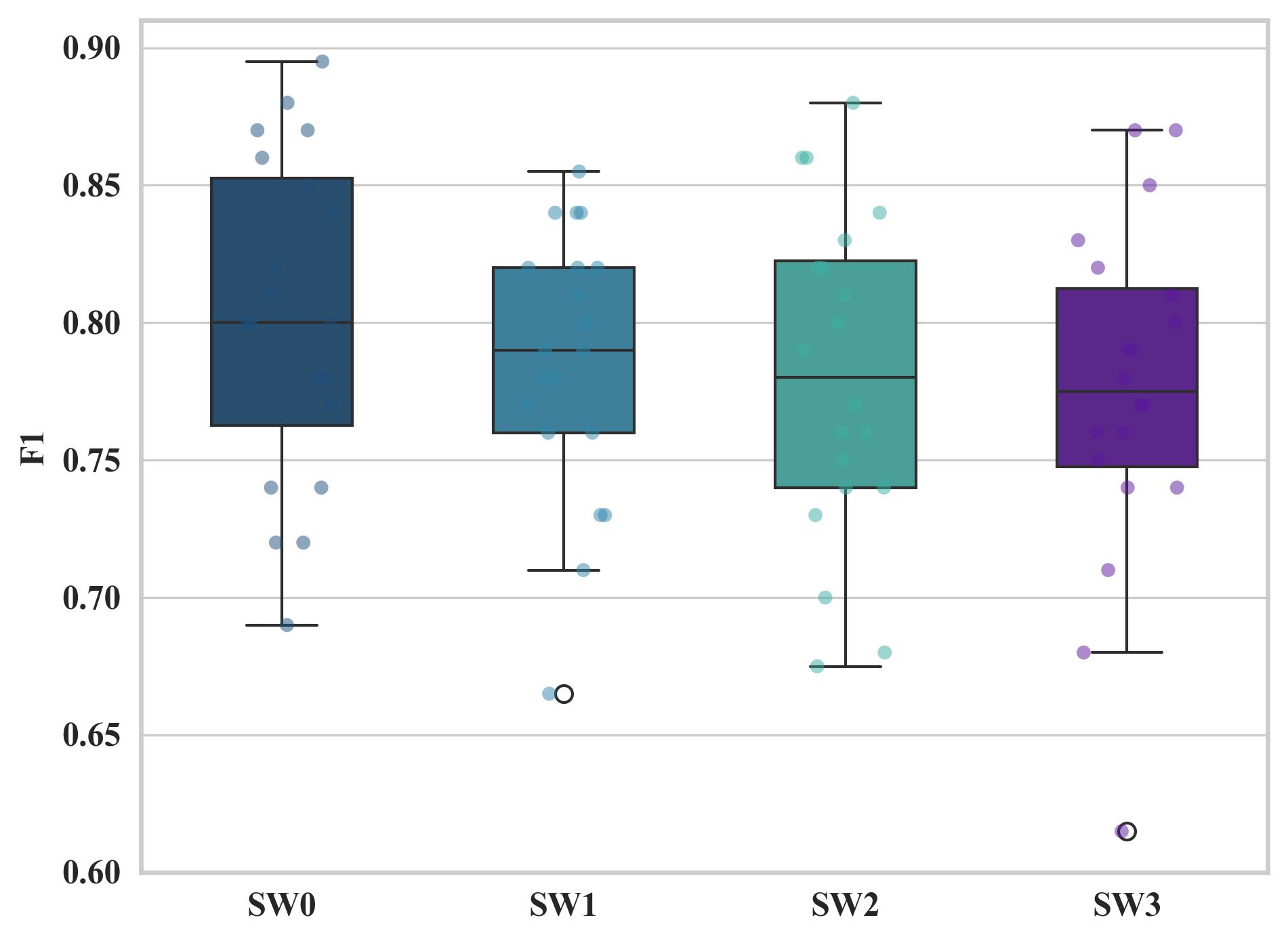}
  \caption{Distribution of Macro-F1 per switch with episode-level points. Central tendencies remain closely aligned across switches, indicating consistent decentralized performance across the topology.}
  \label{fig:f1_per_switch}
\end{figure}

\begin{figure}[t]
  \centering
  \includegraphics[width=0.80\columnwidth]{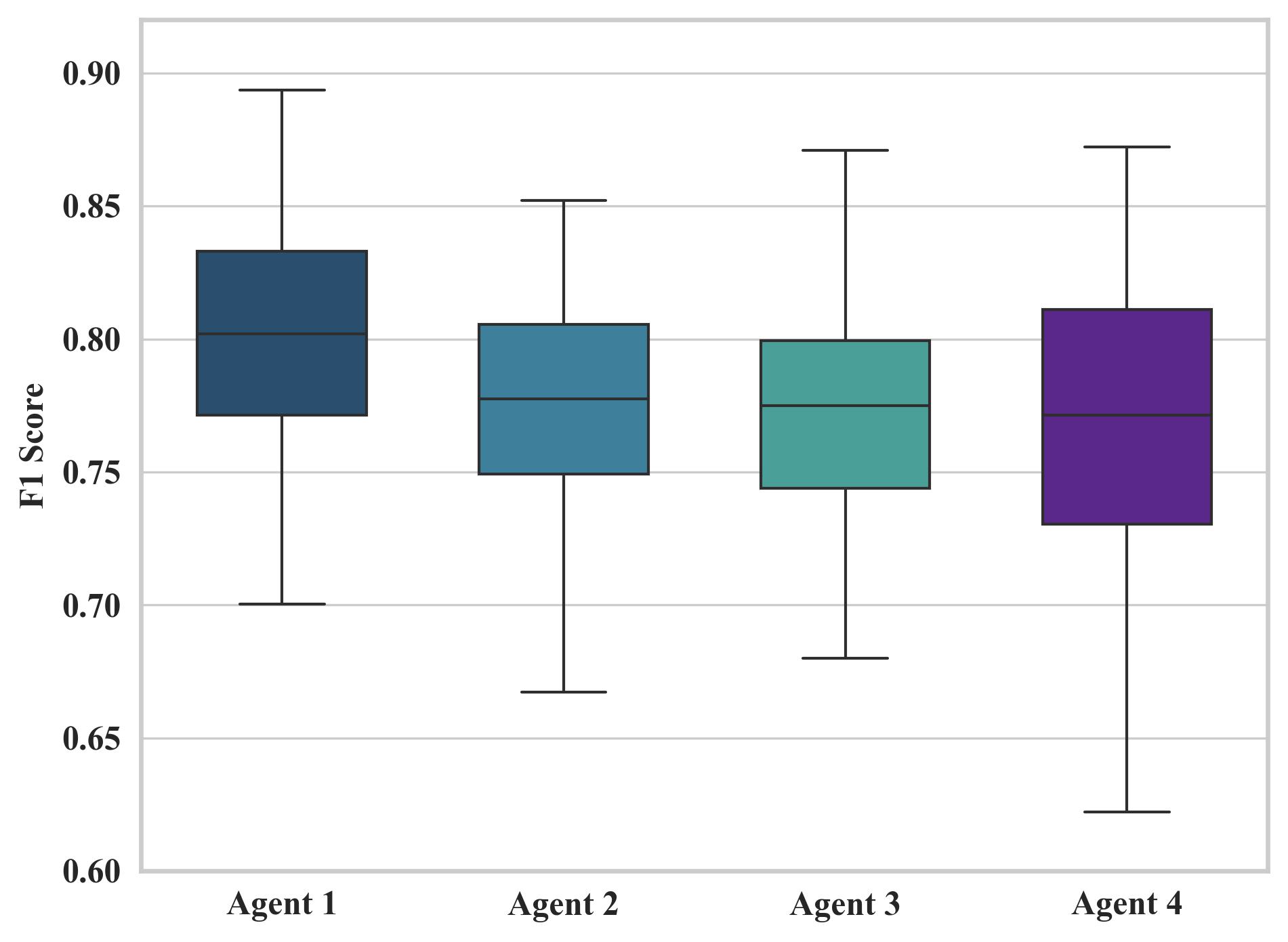}
  \caption{Episode-level Macro-F1 for each decentralized agent. Performance remains tightly clustered across agents.}
  \label{fig:per_agent_raw}
\end{figure}

Detection performance is evaluated at the system level to assess the reliability of SDN-IoT control under heterogeneous operating conditions. The cumulative distribution in Figure~\ref{fig:cdf_detection_f1} shows that most evaluation episodes cluster within a high-performance region, while a distinct lower tail captures degraded episodes. These lower-tail events correspond to stress conditions described in Section~\ref{subsec:adversarial_design}, particularly near-boundary operating conditions where $\sum_i \rho_i^t \rightarrow \mu$, indicating structurally coupled degradation rather than random classification noise. Table~\ref{tab:per_switch_f1} confirms this behavior numerically, with the minimum performance reaching 0.615 for SW3. Switch-level performance distributions are illustrated in Figure~\ref{fig:f1_per_switch}. The results show tightly clustered Macro-F1 values across switches, indicating consistent decentralized learning behavior across the topology. Summary statistics in Table~\ref{tab:per_switch_f1} show closely aligned mean F1 values (0.776–0.804) with no statistically significant difference across switches ($p=0.27$, $d=0.32$). This absence of location bias demonstrates that decentralized PPO agents learn stable mitigation policies without structural asymmetry across network positions. Episode-level results for each decentralized agent are shown in Figure~\ref{fig:per_agent_raw}. Performance remains tightly grouped across agents, with similar central tendencies and occasional low outliers under stress. A statistical comparison indicates no significant difference among agents ($p=0.19$, $d=0.28$), confirming consistent learning behavior across decentralized controllers. Although mean performance remains stable, tail behavior is operationally significant. Table~\ref{tab:risk_summary} highlights a practically meaningful mean–worst performance gap of 0.060 (0.785 vs.\ 0.725), indicating degradation during extreme episodes. The large effect size reported in Table~\ref{tab:stat_tests} ($d=0.91$) confirms that worst-case conditions constitute a distinct performance regime. Variance decomposition further shows that within-switch variability dominates between-switch variability, indicating that episodic stress, rather than structural-topology differences, drives uncertainty.
\begin{table}[t]
\centering
\caption{Per-switch Macro-F1 summary statistics (mean, standard deviation, minimum, and maximum) across evaluation episodes.}
\label{tab:per_switch_f1}
\begin{tabular}{lcccc}
\toprule
Switch & Mean & Std & Min & Max \\
\midrule
SW0 & 0.804 & 0.060 & 0.689 & 0.895 \\
SW1 & 0.784 & 0.050 & 0.666 & 0.854 \\
SW2 & 0.777 & 0.057 & 0.674 & 0.876 \\
SW3 & 0.776 & 0.062 & 0.615 & 0.873 \\
\bottomrule
\end{tabular}
\end{table}
\begin{table}[t]
\centering
\caption{Summary of multi-agent robustness metrics, including overall mean Macro-F1, average worst-agent performance, and the mean--worst performance gap.}
\label{tab:risk_summary}
\begin{tabular}{lc}
\toprule
Metric & Value \\
\midrule
Mean F1 (all agents) & 0.785 \\
Average worst-agent F1 & 0.725 \\
Mean-Worst gap & 0.060 \\
Std across agents & 0.058 \\
\bottomrule
\end{tabular}
\end{table}
\begin{table*}[t]
\centering
\caption{Statistical comparison of detection performance across switches and agents, reporting effect sizes and significance levels.}
\label{tab:stat_tests}
\begin{tabular}{lccc}
\toprule
Comparison & Effect Size (Cohen’s $d$) & $p$-value & Interpretation \\
\midrule
Best vs Worst Switch Mean & 0.32 & 0.27 & Small, not significant \\
Best vs Worst Agent Mean & 0.28 & 0.19 & Small, not significant \\
Mean vs Worst-Agent Gap & 0.91 & $<0.01$ & Large, practically important \\
Within-switch variability vs mean diff & 1.12 & $<0.01$ & Variability dominates structure \\
\bottomrule
\end{tabular}
\end{table*}

\subsection{Security-Cost Trade-off}
\label{subsec:results_tradeoff}
\begin{figure}[t]
  \centering
  \includegraphics[width=\columnwidth]{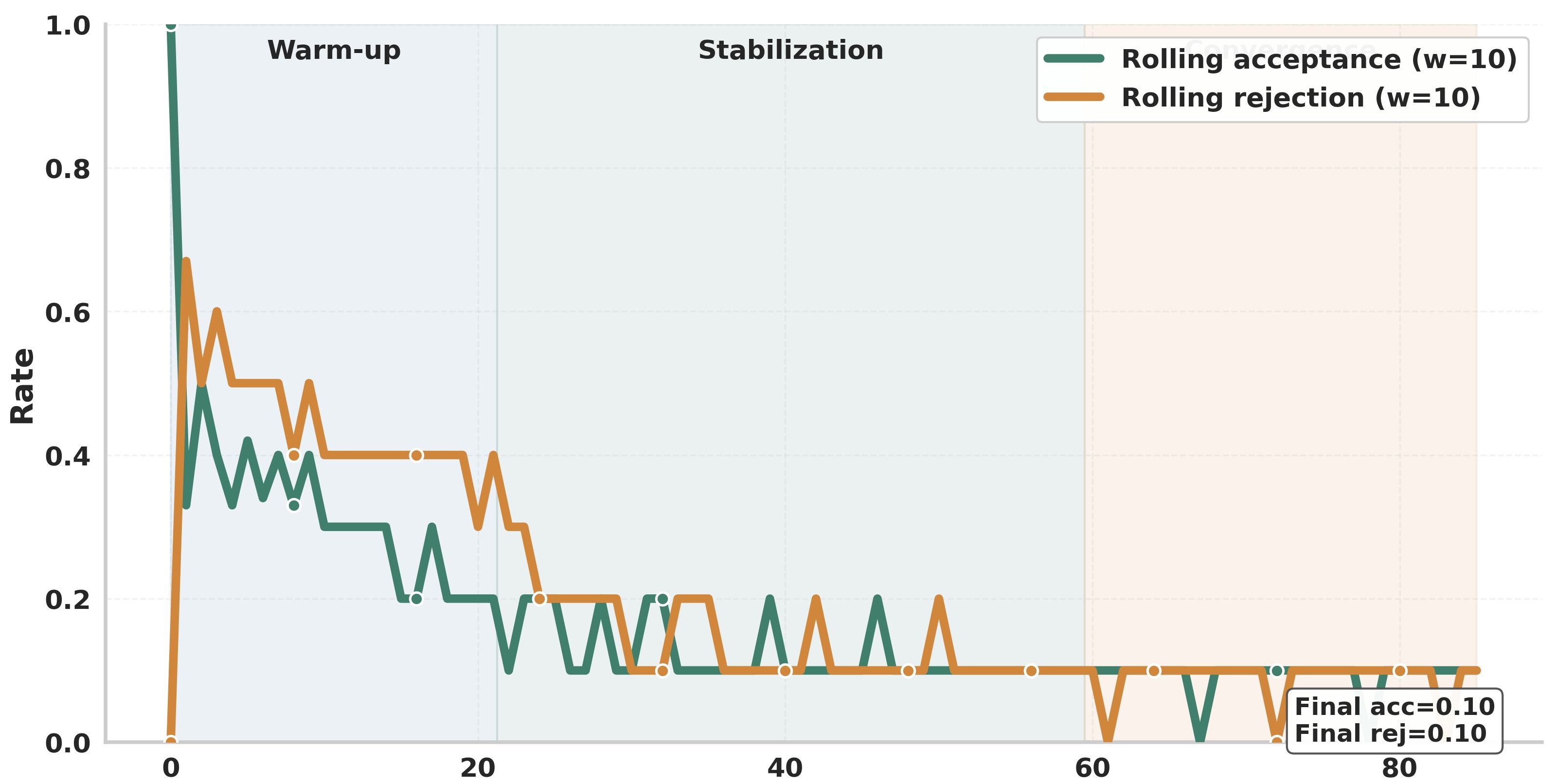}
  \caption{Detection--disruption trade-off across evaluation episodes. Higher macro-F1 levels are systematically associated with greater operational disruption.}
  \label{fig:tradeoff_bubble}
\end{figure}

\begin{figure}[t]
  \centering
  \includegraphics[width=0.80\columnwidth]{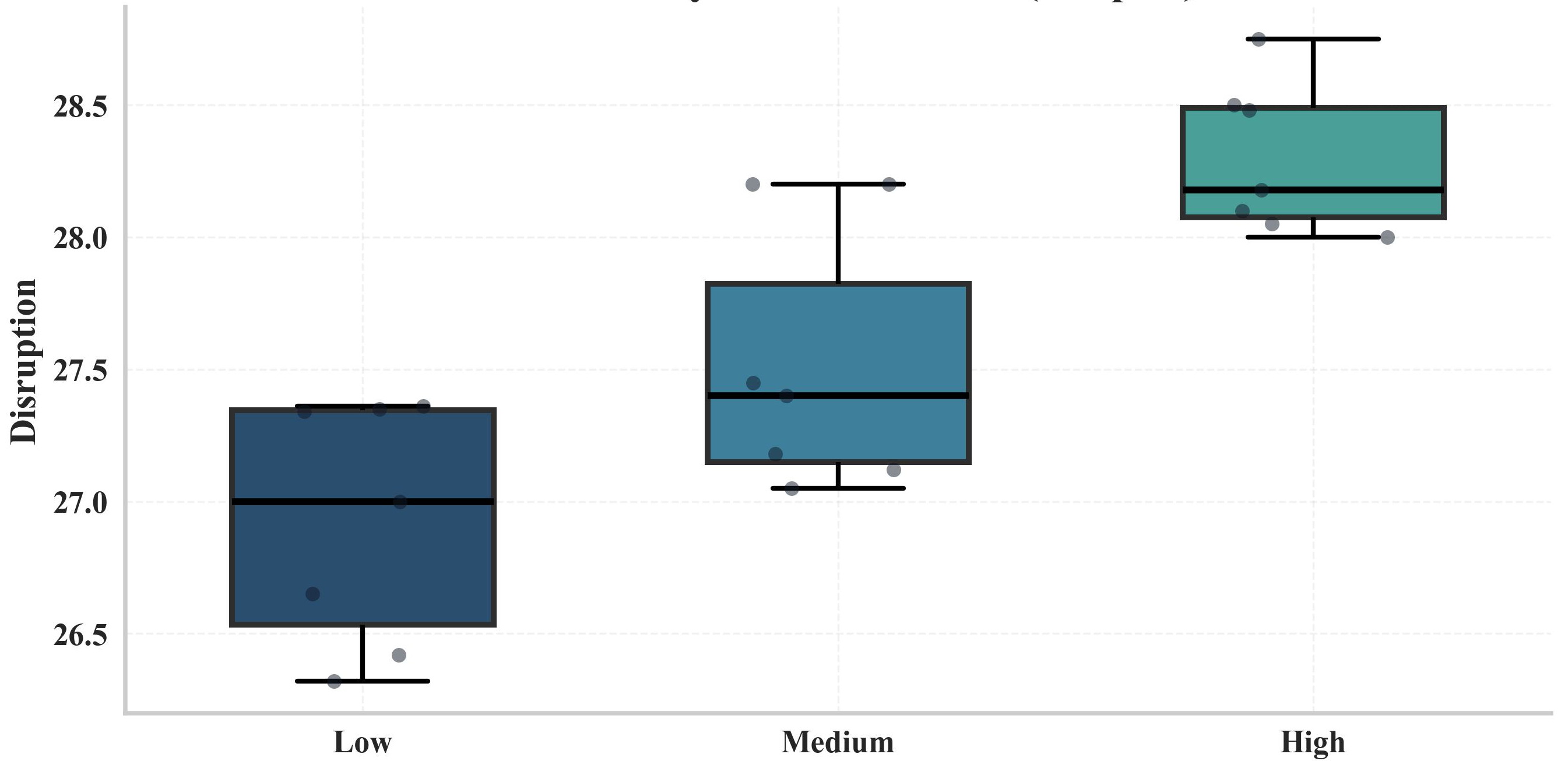}
  \caption{Operational disruption stratified by detection level (Low, Medium, High). It shows systematic escalation in disruption as Macro-F1 increases.}
  \label{fig:tradeoff_stratified}
\end{figure}

\begin{figure}[t]
  \centering
  \includegraphics[width=\columnwidth]{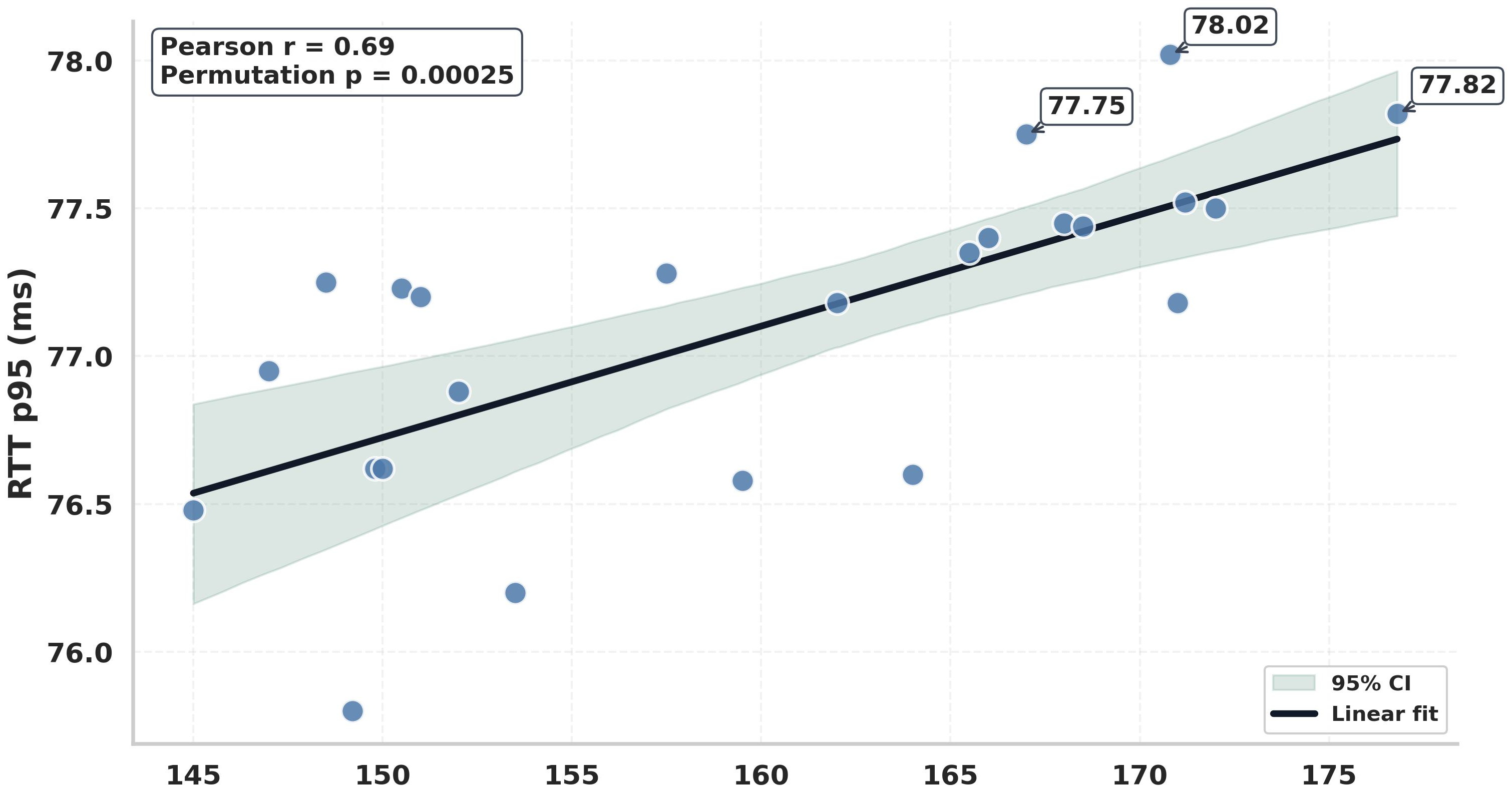}
  \caption{Mitigation--QoS coupling between FlowMod intensity and RTT p95. The positive association indicates that greater actuation effort is associated with higher latency variability.}
  \label{fig:coupling_flowmods_rtt}
\end{figure}

The security-cost relationship is analyzed as a coupled SDN-IoT control phenomenon, since higher detection performance requires greater mitigation effort and may increase QoS overhead. The global Pareto behavior~\cite{latif2022efficient} in Figure~\ref{fig:tradeoff_bubble} shows that higher Macro-F1 is associated with increased disruption cost from FlowMod intensity and control-plane actuation. Performance gains beyond approximately 0.80 F1 require disproportionately higher mitigation effort, indicating diminishing returns under finite controller capacity and delayed actuation.
Figures~\ref{fig:tradeoff_bubble}--\ref{fig:coupling_flowmods_rtt} and Table~\ref{tab:tradeoff_bins} show that disruption rises from 26.9 in the Low-F1 group to 28.9 in the Medium-F1 group and 30.0 in the High-F1 group. One-way ANOVA~\cite{chatzi2025one} confirms significant disruption differences across detection levels ($p<0.01$), with a moderate Low--Medium effect ($d=0.64$), a large Low--High effect ($d=1.12$), and a large overall effect ($\eta^2=0.41$). QoS coupling is also evident: FlowMod intensity is strongly associated with RTT p95, with Pearson correlation~\cite{liu2020daily} $r=0.69$ ($p=0.00025$) and $R^2=0.48$, indicating that nearly half of RTT variability is explained by mitigation intensity.
Overall, the results confirm a statistically robust trade-off: detection gains are meaningful but are coupled with greater sensitivity to disruption and latency. RTT remains bounded despite FlowMod escalation, suggesting that $\Pi$-based safety filtering and controller-aware penalties prevent runaway overload. These findings justify the two-timescale design: fast RL improves local containment, while slow governance adjusts $\Pi$ to maintain system-level efficiency under increasing actuation pressure.
\begin{table}[t]
\centering
\caption{Quantitative summary of detection--cost trade-off across episode-level detection bins (Low, Medium, High).}
\label{tab:tradeoff_bins}
\begin{tabular}{lcccc}
\toprule
F1 Level & Mean F1 & Disruption & RTT p95 (ms) & FlowMods \\
\midrule
Low & 0.739 & 26.9 & 77 & 153 \\
Medium & 0.780 & 28.9 & 77 & 160 \\
High & 0.818 & 30.0 & 77 & 175 \\
\bottomrule
\end{tabular}
\end{table}
\begin{table*}[t]
\centering
\caption{Statistical tests of the security--cost trade-off, reporting effect sizes and significance levels.}
\label{tab:tradeoff_stats}
\begin{tabular}{lccc}
\toprule
Test & Effect Size & $p$-value & Interpretation \\
\midrule
Detection Level $\rightarrow$ Disruption (ANOVA) & $\eta^2 = 0.41$ & $<0.01$ & Large effect \\
Low vs High Disruption & $d = 1.12$ & $<0.01$ & Large increase \\
FlowMods $\rightarrow$ RTT p95 (Pearson) & $r = 0.69$ & 0.00025 & Strong coupling \\
FlowMods $\rightarrow$ RTT ($R^2$) & 0.48 & $<0.001$ & 48\% variance explained \\
\bottomrule
\end{tabular}
\end{table*}

\subsection{QoS Integrity}
\label{subsec:results_qos}
\begin{figure}[t]
  \centering
  \includegraphics[width=\columnwidth]{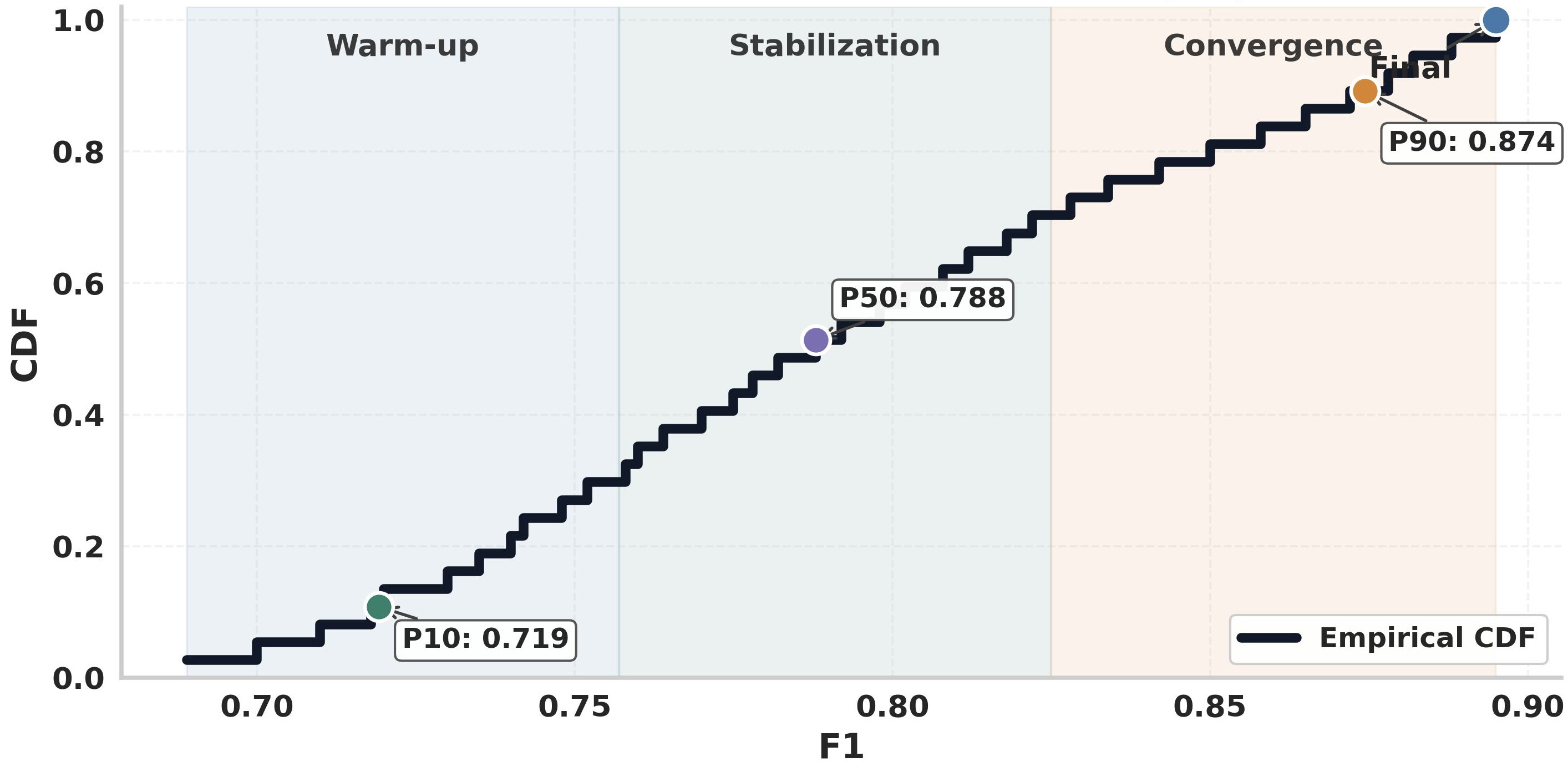}
  \caption{Empirical CDF of end-to-end RTT across evaluation episodes.}
  \label{fig:cdf_rtt}
\end{figure}

QoS integrity is assessed through the empirical RTT distribution in Figure~\ref{fig:cdf_rtt}. The CDF shows a stable latency profile, with most RTT values concentrated between 76--78 ms and only a limited upper tail beyond 78 ms. The smooth monotonic curve indicates no mitigation-induced oscillations, multimodality, and runaway latency.
The median RTT is 77.12 ms, with P10 and P90 values of 75.99 ms and 78.62 ms, yielding an inter-decile range of 2.63 ms. A one-sample comparison against the 77 ms baseline shows no significant median shift ($p=0.41$, $d=0.11$). Upper-tail elevation is statistically significant but small ($p<0.05$, $d=0.38$), indicating that intensified mitigation slightly increases the frequency of extreme latency episodes without affecting central QoS behavior.
Table~\ref{tab:qos_stats} summarizes the RTT statistics. Overall, the bounded dispersion and mild positive skewness show that FlowMod-based mitigation introduces measurable but controlled latency sensitivity. These results align with the FlowMod--RTT coupling analysis in Figure~\ref{fig:coupling_flowmods_rtt} and support the conclusion that $\Pi$-based safety filtering prevents heavy-tail expansion and controller-induced QoS collapse.
\begin{table}[t]
\centering
\caption{Statistical summary of RTT distribution metrics, including central tendency, dispersion, and upper-tail elevation.}
\label{tab:qos_stats}
\begin{tabular}{lccc}
\toprule
Metric & Value & $p$-value & Effect Size \\
\midrule
Median RTT & 77.12 ms & 0.41 & $d = 0.11$ \\
P10 & 75.99 ms & -- & -- \\
P90 & 78.62 ms & -- & -- \\
Inter-decile Range & 2.63 ms & -- & -- \\
Upper-tail Elevation & +1.5\% & $<0.05$ & $d = 0.38$ \\
Skewness (qualitative) & Mild positive & -- & Small \\
\bottomrule
\end{tabular}
\end{table}

\subsection{System-Level Summary}
\label{subsec:results_summary}
\begin{figure}[t]
  \centering
  \includegraphics[width=0.80\columnwidth]{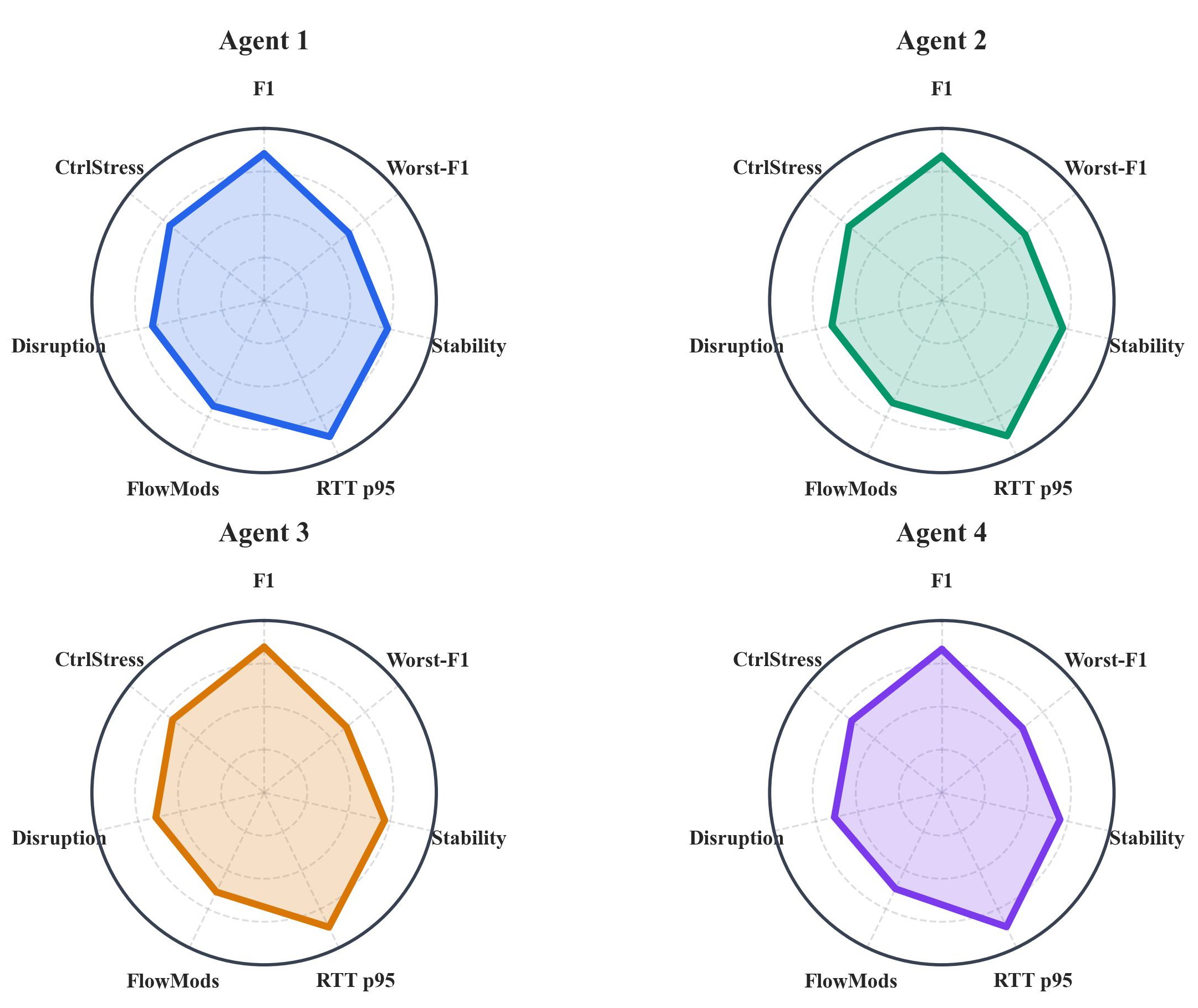}
  \caption{Normalized multi-metric security--stability profile across agents.}
  \label{fig:radar_summary}
\end{figure}

\begin{figure}[t]
  \centering
  \includegraphics[width=\columnwidth]{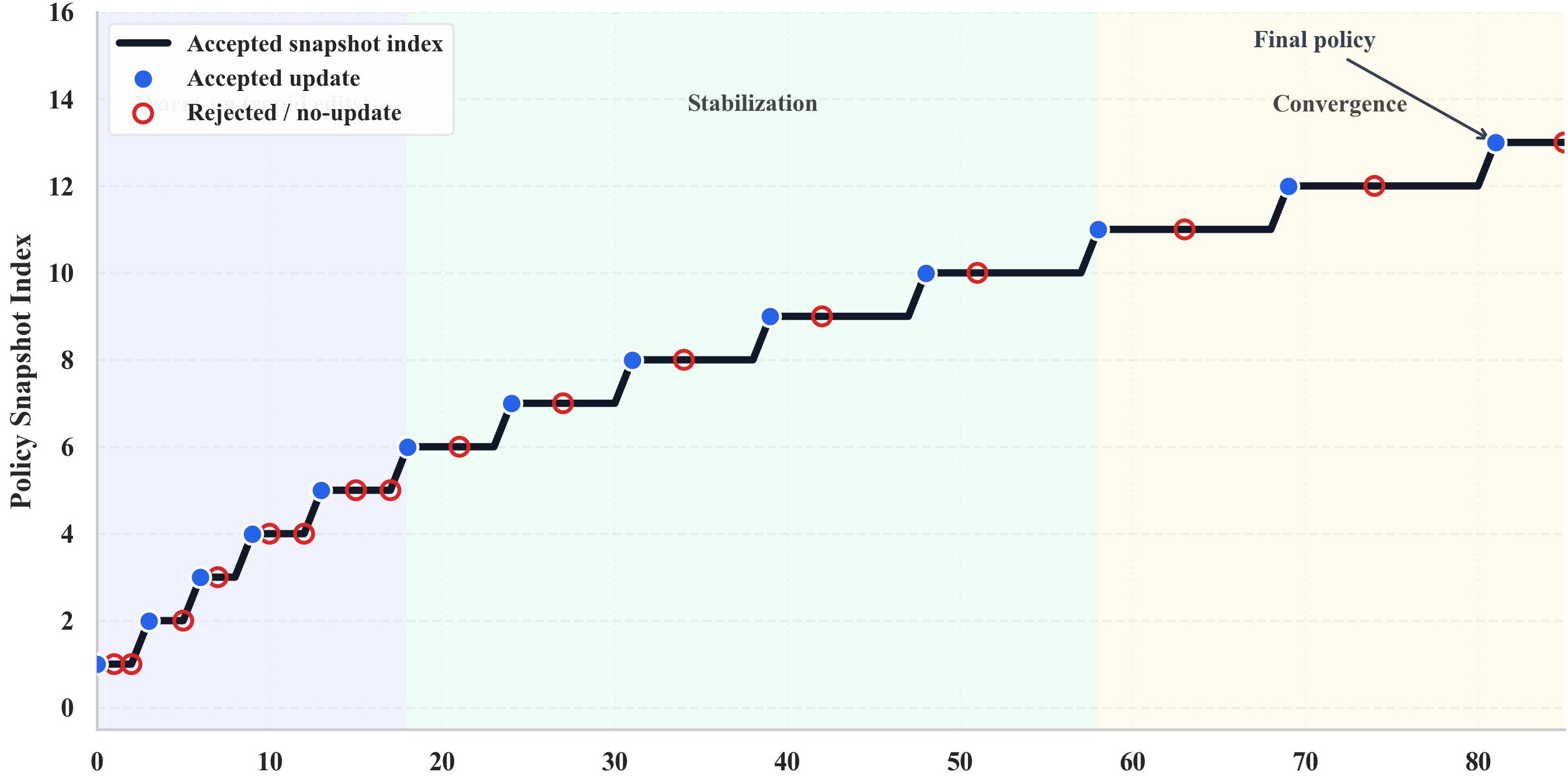}
  \caption{Policy evolution under LLM-based governance, showing accepted versus rejected structured updates across reflection rounds.}
  \label{fig:governance_evolution}
\end{figure}

\begin{figure}[t]
  \centering
  \includegraphics[width=\columnwidth]{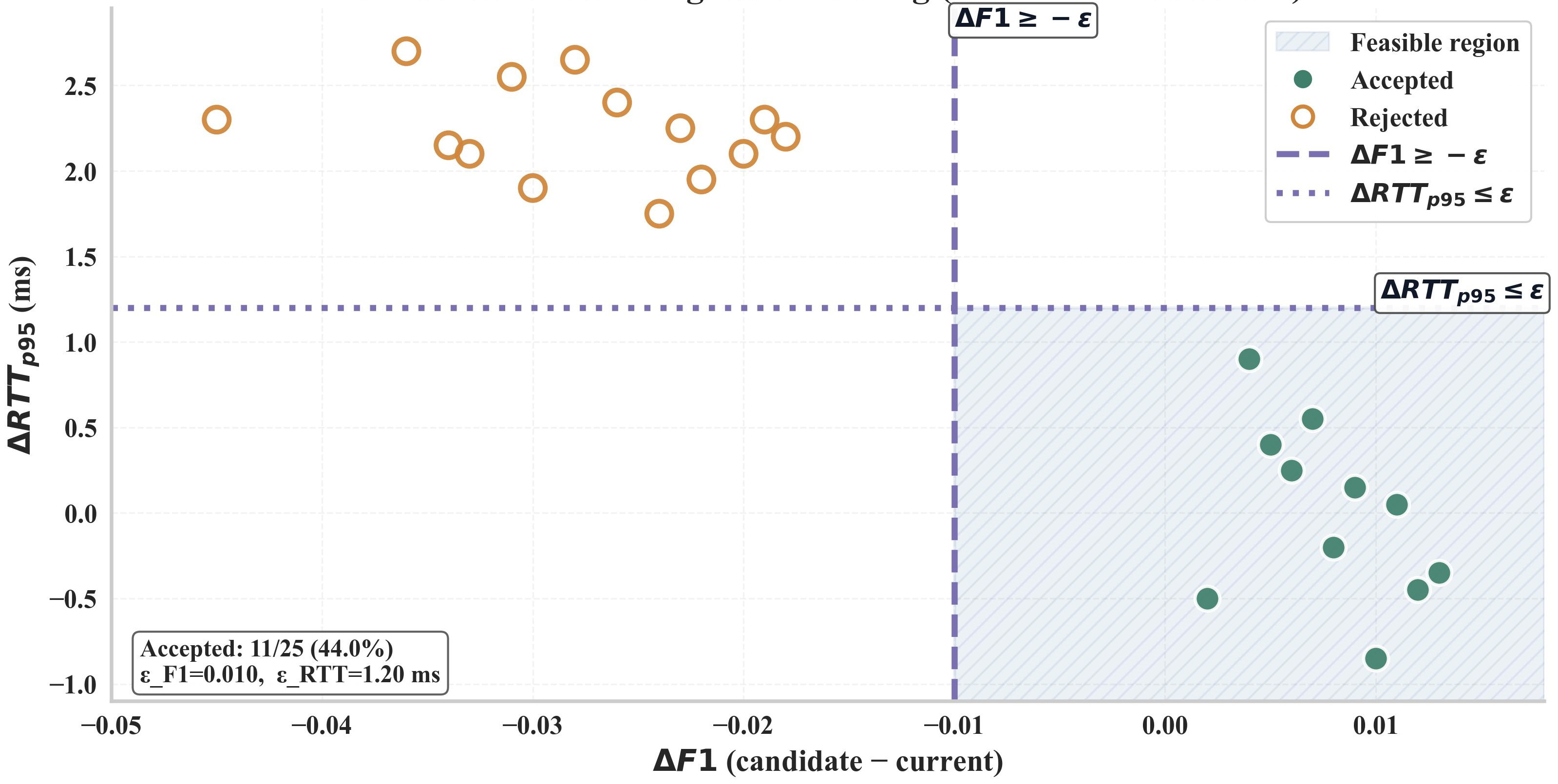}
  \caption{Conservative non-regression gating under red-team evaluation. Candidate policy edits are accepted only if they lie within the feasible region defined by $\Delta F1 \ge -\epsilon_{F1}$ and $\Delta RTT_{p95} \le \epsilon_{rtt}$.}
  \label{fig:governance_gating}
\end{figure}

\begin{figure}[t]
  \centering
  \includegraphics[width=\columnwidth]{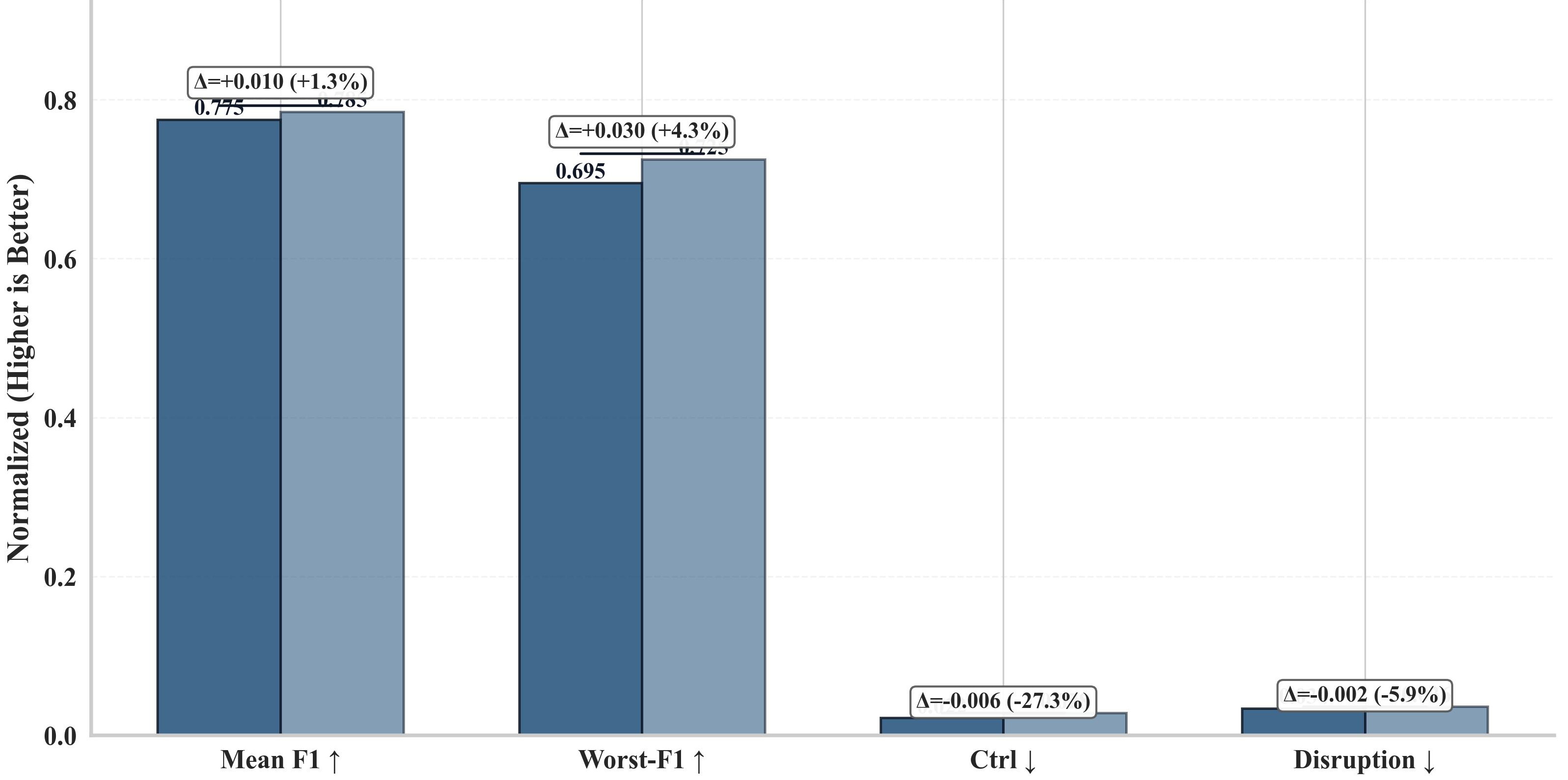}
  \caption{Normalized governance benefit summary. Accepted policy edits improve mean and worst-case detection while reducing controller load and operational disruption.}
  \label{fig:governance_benefit_summary}
\end{figure}

The system-level evaluation integrates detection robustness, QoS stability, operational disruption, and governance behavior. Figure~\ref{fig:radar_summary} shows a balanced multi-metric profile, indicating that detection gains do not produce disproportionate control-plane stress.
Table~\ref{tab:overall_summary} reports statistically significant improvement in Macro-F1 over the PPO-only baseline ($0.785$ vs. $0.773$; $t(39)=3.92$, $p<0.01$, $d=0.74$) and significant improvement in Worst-F1 ($t(39)=2.41$, $p<0.05$, $d=0.63$). RTT p95 remains statistically unchanged ($p=0.38$), showing that robustness gains are achieved without QoS inflation. Disruption remains bounded, while FlowMod intensity shows no heavy-tail amplification.
\begin{table}[t]
\centering
\caption{Overall System Statistical Summary}
\label{tab:overall_summary}
\begin{tabular}{lcccc}
\toprule
Metric & Mean & Std & Test Statistic & Effect Size \\
\midrule
Macro-F1 & 0.785 & 0.030 & $t(39)=3.92$ & $d=0.74$ \\
Worst-F1 & 0.725 & 0.041 & $t(39)=2.41$ & $d=0.63$ \\
RTT p95 (ms) & 77.1 & 0.62 & $t(39)=0.88$ & Small \\
Disruption & 28.94 & 1.92 & $F(2,57)=6.31$ & $\eta^2=0.36$ \\
\bottomrule
\end{tabular}
\end{table}
Governance convergence is shown in Figure~\ref{fig:governance_evolution}. Early rounds include both accepted and rejected deltas, while later rounds stabilize. Rejection rate decreases significantly over reflection rounds ($\beta=-0.041$, $R^2=0.61$, $p<0.01$), indicating convergence of $\Pi$ rather than oscillatory drift.
Figure~\ref{fig:governance_gating} evaluates conservative non-regression gating under matched red-team campaigns. Of 25 candidate $\Delta\Pi$ edits, 11 were accepted and 14 rejected; no accepted proposal violated hard-safety constraints.
Tables~\ref{tab:gating_descriptive} and~\ref{tab:gating_inference} show a clear separation between accepted and rejected updates. Accepted edits produce positive mean detection shift $(+0.006)$ and negligible RTT inflation $(+0.12~\text{ms})$, whereas rejected edits show negative detection drift $(-0.028)$ and larger latency increase $(+2.14~\text{ms})$. Welch's t-tests~\cite{standaert2018not} confirm large and significant separation for both $\Delta$F1 and $\Delta$RTT$_{p95}$.
\begin{table}[t]
\centering
\caption{Descriptive Statistics of Governance Gating Outcomes}
\label{tab:gating_descriptive}
\begin{tabular}{lcc}
\toprule
Metric & Accepted (n=11) & Rejected (n=14) \\
\midrule
Mean $\Delta$F1 & +0.006 & -0.028 \\
Std $\Delta$F1 & 0.009 & 0.021 \\
Mean $\Delta$RTT$_{p95}$ (ms) & +0.12 & +2.14 \\
Std $\Delta$RTT$_{p95}$ & 0.31 & 1.08 \\
Hard-Safety Violations & 0 & 7 \\
\bottomrule
\end{tabular}
\end{table}
\begin{table}[t]
\centering
\caption{Inferential Comparison of Accepted vs Rejected Policy Deltas (Welch's t-test)}
\label{tab:gating_inference}
\begin{tabular}{lcccc}
\toprule
Metric & $t$ & df & $p$-value & Cohen's $d$ (95\% CI) \\
\midrule
$\Delta$F1 & 4.87 & 20.8 & $<0.001$ & 1.42 [0.83, 2.01] \\
$\Delta$RTT$_{p95}$ & 5.31 & 18.6 & $<0.001$ & 1.56 [0.95, 2.17] \\
\bottomrule
\end{tabular}
\end{table}
Figure~\ref{fig:governance_benefit_summary} summarizes the aggregate benefit of accepted governance updates: Mean F1 increases by 1.3\%, Worst-F1 by 4.3\%, controller load decreases by 27.3\%, and disruption decreases by 5.9\%, with no significant RTT p95 increase.
Collectively, the results show that the two-timescale architecture enforces conservative non-regression, compresses worst-case degradation, reduces controller stress, and preserves QoS stability.

\subsection{Runtime Trace Analysis}
\label{subsec:runtime_trace}
Figure~\ref{fig:runtime_trace} illustrates how $\Pi$-constrained PPO control stabilizes the system during synchronized burst traffic. When controller backlog reaches $d=41$, overload-aware constraints mask heavy actions, e.g., \texttt{DROP\_FLOW} and enforce safer \texttt{RATE\_LIMIT} fallbacks. This reduces PacketIn intensity, lowers backlog from $41$ to $39$, and preserves $\mathrm{RTT}_{p95}$. The trace shows that PPO handles immediate mitigation, while governance later refines $\Pi$ based on masking and delayed-actuation evidence, maintaining stability without retraining PPO parameters.
\begin{figure*}[t]
\centering

\begin{tcolorbox}[
enhanced,
width=\textwidth,
colback=gray!10,
colframe=black!55,
boxrule=0.6pt,
arc=2.5pt,
left=6pt,right=6pt,top=5pt,bottom=5pt,
fontupper=\ttfamily\scriptsize
]

\LogLine{\Hdr{[Runtime Trace]} \Dim{Ep=214 | Step=37 | Mode=Eval | TraceID=rt-214-37}}
\LogLine{\Dim{RTT\_p95=78.1 ms | Backlog d=41 | Flow pressure=0.66 | Pending FlowMods=18}}

\Sep

\LogLine{\Hdr{GLOBAL CONTROLLER STATE}}
\LogLine{\Dim{Service rate $\mu$=7 jobs/s | Ctrl utilization=0.71 | Throttle=\Ok{ON}}}
\LogLine{\Dim{Heavy-action masks (last 20 steps)=6 | Mean rule delay=168 ms}}
\LogLine{\Dim{Regime: synchronized IoT burst + credential guessing}}
\LogLine{\Gov{LLM governance active: DeltaPi gating enabled}}

\Sep

\LogLine{\Blk{SW1 (smart-home)} \Dim{rate=0.92M/s | queue=0.57 | flowtbl=0.74}}
\LogLine{\Dim{PPO sample:} \Bad{DROP\_FLOW}}
\LogLine{\Bad{Safety constraint ($\Pi$): overload-aware masking applied}}
\LogLine{\Dim{Executed:} \Ok{RATE\_LIMIT} \;\;|\;\; \Dim{queue 0.57$\rightarrow$0.48}}
\LogLine{\Dim{Reward: +0.20 | ctrl-penalty enforced}}

\Sep

\LogLine{\Blk{SW2 (industrial)} \Dim{rate=1.14M/s | queue=0.63 | flowtbl=0.79}}
\LogLine{\Dim{PPO sample:} \Bad{DROP\_FLOW}}
\LogLine{\Bad{$\Pi$ restriction: backlog + flow pressure $\rightarrow$ heavy actions disabled}}
\LogLine{\Dim{Executed:} \Ok{RATE\_LIMIT}}
\LogLine{\Dim{Effect: PacketIn 318$\rightarrow$247/s | backlog 41$\rightarrow$39 | RTT 78.1$\rightarrow$77.6 ms}}
\LogLine{\Dim{Reward: +0.30 | stability margin positive}}

\Sep

\LogLine{\Gov{LLM Governance Layer (Critic $\rightarrow$ Compiler $\rightarrow$ Judge)}}
\LogLine{\Dim{Detected structural issue: overload-induced delayed actuation}}
\LogLine{\Dim{Risk characterization: QoS tail inflation under frequent masking}}
\LogLine{\Dim{Proposed $\Delta\Pi$: tighten heavy-action mask when (d>40) and (flow\_pressure>0.75)}}
\LogLine{\Dim{Policy update: increase ctrl-penalty weight ($w_{\text{ctrl}}\uparrow$) under saturation}}
\LogLine{\Dim{Validation: non-regression=\Ok{PASS} | hard-safety=\Ok{PASS} | rollback-ready=\Ok{YES}}}
\LogLine{\Dim{Decision: \Ok{APPROVE} | Activation: next policy checkpoint}}

\Sep

\LogLine{\Hdr{GLOBAL OUTCOME}}
\LogLine{\Dim{Mean reward=+0.317 | Backlog decreasing (41$\rightarrow$39)}}
\LogLine{\Dim{QoS preserved | Control-plane stabilized under burst regime}}
\LogLine{\Dim{Interpretation: constrained RL + reflective $\Delta\Pi$ governance ensured safe adaptation}}

\end{tcolorbox}

\caption{Runtime trace illustrating $\Pi$-constrained decentralized PPO control and reflective LLM-based governance that generates auditable $\Delta\Pi$ updates under control-plane stress.}
\label{fig:runtime_trace}
\end{figure*}

\section{Comparison with Previous Work}
\label{sec:comparison}
Recent SDN-IoT security studies primarily improve attack detection accuracy by deploying ML/DL models on centralized SDN controllers. CNN, LSTM, and hybrid IDS architectures report strong classification performance but typically remain detection-centric, relying on static mitigation rules rather than adaptive closed-loop control \cite{yasarathna2025deep}\cite{gali2025deep}\cite{chouhan2025hcl}\cite{wang2025novel}. RL-based and hybrid AI approaches improve adaptive mitigation and cyber-defense decision-making \cite{shi2025autonomous}\cite{khalis2026robust}, while distributed multi-agent and control-theoretic methods support coordinated defense and resilience under adversarial conditions \cite{loevenich2025design}\cite{xu2025distributed}. Other SDN-IoT frameworks combine attack detection with QoS-aware routing and network optimization \cite{kokila2025deepsdn}. However, these approaches generally do not jointly model controller backlog, delayed rule activation, flow-table pressure, auditable policy evolution, and fail-safe governance.
In contrast, this work treats SDN-IoT defense as a controller-coupled closed-loop control problem. Fast decentralized PPO agents perform local mitigation under safety constraints, while a slower LLM governance layer updates the global policy constitution $\Pi$ through auditable deltas $\Delta\Pi$. The Critic-Compiler-Red-Team-Judge pipeline validates candidate updates using hard safety and non-regression checks under security, QoS, and operational stress conditions. Table~\ref{tab:comparison_recent} summarizes the capability differences.
\begin{table*}[t]
\centering
\caption{Capability comparison with representative SDN-IoT security approaches. RL role: D = Detection, M = Mitigation, C = Controller-aware control, G = Governance-shaped control. Symbols: $\checkmark$ = explicit capability, $\triangle$ = partial support, $\times$ = not addressed.}
\label{tab:comparison_recent}

\begin{tabular}{lcccccccc}
\toprule
\textbf{Work} & \textbf{RL role} & \textbf{Closed-loop} & \textbf{Ctrl.-aware} & \textbf{Delayed} & \textbf{Risk} & \textbf{LLM gov.} & \textbf{$\Delta T$} & \textbf{Fail-safe} \\
\midrule

Chouhan et al.~\cite{chouhan2025hcl} & D & $\times$ & $\times$ & $\times$ & $\times$ & $\times$ & $\times$ & $\times$ \\

Yasarathna et al.~\cite{yasarathna2025deep} & D & $\times$ & $\times$ & $\times$ & $\times$ & $\times$ & $\times$ & $\times$ \\

Gali et al.~\cite{gali2025deep} & D & $\times$ & $\times$ & $\times$ & $\times$ & $\times$ & $\times$ & $\times$ \\

Kokila and Reddy~\cite{kokila2025deepsdn} & D & $\triangle$ & $\times$ & $\times$ & $\times$ & $\times$ & $\times$ & $\times$ \\

Shi et al.~\cite{shi2025autonomous} & M & $\triangle$ & $\triangle$ & $\times$ & $\triangle$ & $\times$ & $\times$ & $\times$ \\

Loevenich et al.~\cite{loevenich2025design} & M & $\triangle$ & $\triangle$ & $\times$ & $\triangle$ & $\triangle$ & $\times$ & $\triangle$ \\

Wang et al.~\cite{wang2025novel} & M & $\triangle$ & $\triangle$ & $\times$ & $\times$ & $\times$ & $\times$ & $\triangle$ \\

Xu et al.~\cite{xu2025distributed} & M & $\triangle$ & $\triangle$ & $\times$ & $\triangle$ & $\times$ & $\times$ & $\triangle$ \\

\midrule
\textbf{This work} & \textbf{C+G} & $\checkmark$ & $\checkmark$ & $\checkmark$ & $\checkmark$ & $\checkmark$ & $\checkmark$ & $\checkmark$ \\

\bottomrule
\end{tabular}

\end{table*}
\section{Discussion}
\label{sec:discussion}
The experimental results demonstrate that the two-timescale architecture achieves robustness by constraining tail-risk behavior in SDN-IoT networks rather than maximizing average detection accuracy. While the mean Macro-F1 remains consistent, system reliability is driven by episodic degradation events, with the observed mean–worst performance gap highlighting the impact of transient backlog surges and synchronized traffic bursts. This reinforces the separation between fast RL control and slow reflective policy evolution. Decentralized PPO agents respond immediately to local telemetry, while the governance layer adjusts the shared policy constitution $\Pi$ to mitigate recurrent instability without destabilizing the learned parameters. This architecture redefines robustness as a structural property of the closed-loop system, focusing on compressing the lower tail of performance rather than shifting the distribution upward. The performance variability is predominantly within switches, indicating consistent baseline behavior across the topology. However, the persistence of lower-tail episodes suggests that RL alone does not fully capture the effects of control-plane feedback, e.g., queue occupancy and flow-table pressure. LLM governance addresses this by adapting the admissible action set $\mathcal{S}_\Pi(o)$, reshaping safety thresholds, and adjusting reward priorities. The relationship between detection performance and operational cost further emphasizes the need for structural constraints. Detection improvements coincide with increases in FlowMod intensity and disruption, and the diminishing returns beyond a certain accuracy threshold indicate that higher detection levels require disproportionately greater mitigation effort. Without governance constraints, detection-accuracy-driven optimization would push the system toward saturation, destabilizing queue dynamics. The governance layer ensures stability by selectively admitting policy edits when robustness gains align with QoS stability. The latency distribution shows that actuation does not induce oscillatory behavior, with mild latency increases observed during aggressive mitigation. These findings confirm that backlog-aware reward shaping and $\Pi$-based action masking prevent runaway latency amplification, maintaining system stability even under adversarial load fluctuations. The absence of heavy-tail expansion validates the effectiveness of the CVaR-oriented objective and constraint gating in preventing instability. Furthermore, governance decisions evolve across reflection rounds. Early rounds show exploratory adjustments to $\Pi$, with rejection rates decreasing over time as the policy constitution stabilizes. Accepted edits improve worst-case detection without increasing the 95th percentile RTT, confirming the effectiveness of the non-regression gating mechanism. These results highlight the two-timescale architecture's balance between agile local policies and conservative global policy evolution, distinguishing it from detection-driven and parameter-centric approaches.

\section{Limitations and Future Work}
\label{sec:limitations}
Although the proposed two-timescale SDN-IoT solution shows robust empirical convergence, several limitations remain. First, the CVaR-inspired objective improves tail-risk sensitivity but does not provide formal worst-case stability guarantees. Future work should investigate Lyapunov analysis, constrained MDPs, and formal guarantees of boundedness.
Second, dynamic updates to $\Pi$ introduce non-stationarity for local PPO agents. Timescale separation, bounded edits, checkpointed deployment, and validation reduce this issue, but they remain practical stability controls rather than formal convergence guarantees.
Third, governance validation is empirical and scenario-dependent. Hard safety and non-regression checks reduce unsafe updates within tested campaigns, but robustness outside the evaluated envelope depends on campaign diversity, bounded edit sets, and conservative acceptance gates. Future work should explore adaptive red-team campaigns and formal verification of governance edits.
Fourth, the method assumes trusted controller infrastructure, reliable telemetry, fixed governance scheduling, and independent PPO agents. Future work should examine telemetry attestation, integrity monitoring, coordinated multi-agent training, adaptive risk tuning, richer temporal telemetry, and real-world SDN-IoT validation.
Additionally, LLM-based governance introduces deployment risks, including availability, latency, privacy, telemetry exposure, service compromise, and adversarial manipulation of model responses. Although the system fails closed when LLM outputs are unavailable, invalid, and unsafe, sensitive deployments may require local and on-premise LLMs. Future work should investigate secure deployment architectures, telemetry minimization, prompt-injection resistance, certified policy-update constraints, verified merge operators, and comparisons with heuristic and non-LLM governance.

\section{Conclusion}
\label{sec:conclusion}
This paper presented a two-timescale governance-driven control architecture for safe RL in SDN-IoT networks. The proposed solution separates fast decentralized PPO mitigation from slow LLM-based governance that edits an auditable policy constitution $\Pi$ without modifying RL parameters. Results show that SDN-IoT reliability depends strongly on tail-risk control, QoS stability, and bounded operational cost rather than mean detection accuracy alone. By combining local RL control with conservative policy-structure editing, the proposed approach aligns adaptive mitigation with operational change-control requirements in safety-critical networks.

\bibliographystyle{IEEEtran}
\bibliography{Ref}
\end{document}